\documentclass[iop]{emulateapj}

\shorttitle{Variable stars in G11}
\shortauthors{Contreras Ramos et al.}

\begin{document}

\title{Does the  Oosterhoff Dichotomy Exist in The Andromeda Galaxy?\\
I. The case of G11\altaffilmark{*}}

\author{Rodrigo Contreras Ramos\altaffilmark{1,2,3}, 
Gisella Clementini\altaffilmark{1}, Luciana Federici\altaffilmark{1}, Giuliana Fiorentino\altaffilmark{1},  
Carla Cacciari\altaffilmark{1},  M\'arcio Catelan\altaffilmark{3,8}, Horace A. 
Smith\altaffilmark{4},  Flavio Fusi Pecci\altaffilmark{1}, Marcella 
Marconi\altaffilmark{5}, Barton J. Pritzl\altaffilmark{6}, and Karen Kinemuchi\altaffilmark{7}}

\altaffiltext{1}{INAF, Osservatorio Astronomico di Bologna, Bologna,
Italy; rodrigo.contreras@oabo.inaf.it}
\altaffiltext{2}{Dipartimento di Astronomia, Universit\`a di Bologna, Bologna, Italy}
\altaffiltext{3}{Pontificia Universidad Cat$\rm{\acute{o}}$lica de Chile, Departamento de Astronom\'{\i}a y Astrof\'{\i}sica, 
Santiago, Chile; mcatelan@astro.puc.cl}
\altaffiltext{4}{Department of Physics and Astronomy, Michigan State
University, East 
Lansing, MI 48824, USA; smith@pa.msu.edu}
\altaffiltext{5}{INAF, Osservatorio Astronomico di Capodimonte, Napoli, Italy; marcella@na.astro.it}
\altaffiltext{6}{Department of Physics and Astronomy, 
University of Wisconsin Oshkosh, Oshkosh, WI 54901, USA; pritzlb@uwosh.edu}
\altaffiltext{7}{Apache Point Observatory/New Mexico State University, Sunspot,NM 88349, USA; 
kinemuchi@apo.nmsu.edu}
\altaffiltext{8}{The Milky Way Millennium Nucleus, Av.\ Vicu{\~n}a Mackenna 4860, 782-0436 Macul, Santiago, Chile}
\altaffiltext{*}{Based on data collected with the Wide Field Planetary Camera 2 on board of the
{\it Hubble Space Telescope}.}
%---------------

\begin{abstract}
We present the first evidence that Oosterhoff type II globular
clusters exist in the Andromeda galaxy (M31). On the basis of time--series photometry 
of  the moderately metal--poor ([Fe/H]$\sim -$1.6 dex) M31 globular cluster G11,
obtained with the Wide Field Planetary Camera 2 on board the {\it  Hubble Space
Telescope}, we detected 
and derived periods for 14 RR Lyrae stars, of which five are found to lie inside the cluster tidal radius.
They include three 
fundamental--mode (RRab) and two
first--overtone (RRc) pulsators, with average periods 
$\langle P_{ab} \rangle$ = 0.70 d, and $\langle P_{c} \rangle$ = 0.40 d, 
 respectively. These mean periods  
and the position of the cluster variable stars in the
period-amplitude and period-metallicity diagrams, all suggest that G11 is likely to be an Oosterhoff 
type II globular cluster. This appears to  be in agreement with the 
general behavior 
of Milky Way globular clusters with similar metallicity and 
horizontal branch morphology. 
\end{abstract}


%%%%%%%%%%%%%%%%%%%%%%%%% SEC. 1: INTRO
\section{Introduction}\label{intro}

More than seventy years ago, \cite{oosterhoff39} discovered  that the Milky Way (MW)  
globular clusters (GCs) could be divided into two groups  according 
to the mean period of their  RR Lyrae (RRL) stars. The two groups were later christened according to his name, 
Oosterhoff type I (Oo~I) and Oosterhoff type II (Oo~II; see, e.g., \citealt{preston59}). 
Oo~I clusters contain  
fundamental-mode RRL stars (RRab) with a mean period, 
$\langle P_{ab}\rangle$, close to 0.55 d,  
while Oo~II clusters contain RRab's  with mean 
period close to 0.65 d. 
Differences in the mean period of the first-overtone RRL stars (RRc) in these clusters were also present. 
Oo~I clusters had the RRc mean period, $\langle P_{c}\rangle$, near 
0.32 d, while GCs classified as Oo~II showed
$\langle P_{c}\rangle \sim 0.37$ d. He also noted that the percentage of
{\it c-}type variables differed between the two groups, with  the RRL population of the Oo~I clusters being largely
dominated by {\it ab-}type pulsators ($f_c = N_c/N_{ab+c} \sim 0.17$, where $N_{c}$  and $N_{ab}$
are the numbers of first--overtone and fundamental--mode RRL stars, respectively), 
whereas Oo~II clusters
generally contained a more balanced proportion of $c-$ to $ab-$type
pulsators ($f_c \sim$ 0.44). However, one should be aware that incompleteness 
and selection effects may alter this property \citep{smith95}.
Subsequent studies \citep{oosterhoff44,sawyer44} including new RRL-rich 
GCs soon confirmed the dichotomy.
When metal abundances began to be measured for GCs,  it became clear 
that the two Oo groups showed differences  also in metallicity. 
Oo~I clusters were moderately metal-poor ([Fe/H] $\sim -$1.3 dex), 
while Oo~II clusters were very metal-poor ([Fe/H] $\sim -$2.0 dex). 
Recent studies \citep[see e.g.][for a review]{catelan09}
involving larger samples of GCs, more accurate metallicities, 
better periods for the RRL stars, and higher completeness,  have 
fully confirmed the Oosterhoff dichotomy not only among the Galactic GCs (GGCs), but also for 
field variable stars in the MW halo \citep[see e.g.][]{miceli08}.
On the other hand, extragalactic GCs and field RRL stars in the Local Group dwarf
galaxies generally have $\langle P_{ab} \rangle$ ranging between 
0.58 d and 0.62 d, thus they preferentially occupy the so-called Oosterhoff gap avoided by 
the MW GCs, and are classified as Oosterhoff-Intermediate \citep[Oo-Int;][and references
therein]{catelan09,clementini10}\footnote{Among the bright dwarf spheroidal (dSph) companions of the MW, only Ursa Minor with $\langle P_{ab}\rangle \sim 0.64$ days has Oo~II 
properties, and Sagittarius with $\langle P_{ab}\rangle \sim 0.574$ days  is on the long period tail of the Oo~I group \citep[see e.g.][]{cseresnjes01}.}. This very straightforward observational evidence bears  very significant 
implications on our understanding of the Galactic
halo formation. Indeed, if the Galactic  halo was assembled by capture of ``building blocks" 
resembling the bright dwarf galaxies we see today around the MW, then the properties of their  RRL stars should conform to the Oosterhoff dichotomy we
 observe for the MW field and cluster stars. 
 The absence of an Oosterhoff dichotomy, along with differences in the chemical abundances (e.g.,  \citealt{shetrone01},   \citealt{helmi06}),
 thus seem to rule out the bright dwarf galaxies surrounding the MW  as significant contributors to the assembling of the 
 Galactic halo \citep[e.g.][]{catelan04,catelan09}. 
 
With this paper we turn our interest to the other large spiral in the Local Group, the Andromeda galaxy (M31). %,  to ask the question of whether 
The systematic study of the variable star population in M31 is 
fairly recent.  Indeed, at a distance of 
%780 kpc {\citep[][]{mcconnachie05}, 
752 kpc $\pm$ 27 {\citep[][]{riess12},
at least 4-m telescopes (see, e.g., \citealt{pritchet87}) and preferably 8m-class 
telescopes are needed to resolve and measure individual stars in the field 
of Andromeda from the ground, while only the resolution capabilities of the {\it Hubble Space
Telescope} ({\it HST}) allow to resolve individual stars 
on different evolutionary paths, including the RRL stars, in GCs 
or crowded fields in Andromeda. 
Only five ``classical'' dSph satellites of M31 have been analysed for variability 
so far (And VI: \citealt{pritzl02}; And II: \citealt{pritzl04}; And I \&
III: \citealt{pritzl05};  and And V: \citealt{mancone08})
and, unlike the MW dSphs, all three Oosterhoff types 
(Oo~I, Oo~II, and Oo-Int) were observed. 
As far as the field of M31 is concerned, our current knowledge 
relies on a few relatively recent studies which have
provided a detailed analysis of more than a thousand RRL stars in the M31 field.
Different components of M31 were analysed, 
e.g., halo, disk, stream \citep{brown04,jeffery11} 
including regions close to M32 \citep{sarajedini09,
fiorentino10b,fiorentino12,sarajedini12}, and all these studies agree on  the  M31 field variables having properties consistent with an Oosterhoff type I classification. 
For the study of the Oosterhoff dichotomy, however, cluster RRL stars
are needed.
A pioneering study of the RRL stars in four M31 GCs was presented by \citet{clementini01}, who detected a number of RRL candidates 
in the M31 clusters G11, G33, G64, and G322, based on $F555W$, $F814W$ {\it HST} Wide Field Planetary Camera 2 (WFPC2) archive observations.
However, the number of phase points and the time window of these archive data were too sparse to allow the derivation of the variable stars' 
periods. Thus, we do not know yet 
whether the M31 GCs show the Oosterhoff dichotomy, or whether indeed they
can be placed into an Oosterhoff type at all. 

WFPC2 observing time  to  study the RRL stars in six  M31 GCs
was granted to us  in {\it HST} Cycle 15, under program GO 11081 (PI: G. Clementini). 
We  obtained time-series photometry for  G11, G33, G76, G105, G322, and B514 for which the basic parameters are reported in  Table~\ref{table01}. 
These M31 GCs have 
well--defined Color-Magnitude Diagrams (CMDs) (Rich et al. 2005; Galleti et al. 2009), 
metallicities compatible with the existence of RRL stars, as well as populous horizontal branches (HBs) stretching across the 
instability strip (IS), hence 
providing a good probability of having RRL stars. 
Three of these GCs $-$ G76, G105, G322 $-$ have a metallicity that, in the MW, would place them
in the Oo~I group, whereas the other three $-$ G11, G33, B514 $-$ are more metal-poor and, in the MW, 
would be expected to be Oo~II GCs. 
Fig.~\ref{targets} shows the positions of these six GCs
on a $\sim 2.9 \times 2.9$ deg$^2$ image of M31. They span 
distances in the range of $\sim 45$ arcmin (G76) to  $\sim 4$ degrees (B514) from the center of M31, thus
 allowing us to probe different areas 
of Andromeda, from the galaxy's unperturbed halo to regions close to the 
giant stream.
We are aware that these few clusters may not necessarily be representative of 
the entire 
GC ensemble of M31, still they allowed us to check for the 
first time whether the properties of the RRL stars in the M31 GCs 
are compatible with the Oosterhoff dichotomy,  and whether 
they show the same correlations with HB morphology as seen in the MW.
Noteworthy, the presence of the Oosterhoff dichotomy in the MW GC system was first established on the basis of a study of
the RRL properties in only five MW clusters, of which three turned out 
to be of Oosterhoff type II, and two of Oosterhoff type I. 
A first characterization of the variable stars of  B514  based on this dataset was published  by \citet{clementini09},
who detected and derived periods for 89 RRL stars (82 RRab and 7 RRc pulsators) among 161 candidate variables identified in the cluster. 
The average period of the RR Lyrae variables, and the position in the period-amplitude diagram both suggested for B514 an Oo~I  
classification,  in spite of its low metallicity ([Fe/H]=$-$2.06, \citealt{galleti09}).  A preliminary study of the other  five M31 GCs 
has been presented in \citet{contr10}. 

In this paper we present results from the detailed study of the variable star population in G11 (Bo293) based both on our Cycle 15 observations and on the {\it HST} archival data analyzed by \citet{clementini01}. We also describe in detail the overall strategy we have applied to characterize the variable star population in all the M31 GCs we observed in Cycle 15.
The results for these other  M31 GCs  will be  presented  in two forthcoming papers. 

%%%%%%%%%G11
G11 (Bo293; R.A.(J2000) = {\rm 00$^h$36$^m$20\fs86}, 
decl.(J2000) = {\rm 40$^{\circ}$53$^{\prime}$37\farcs25}), 
 is a relatively massive ($M_V\thickapprox-8.5$ mag) GC,
at a projected distance of $\sim75.75$ arcmin from the center of M31 (Revised Bologna Catalog [RBCV4.0], \citealt{galleti04}\footnote{http://www.bo.astro.it/M31/}).
According to its spatial location the main source of contamination for G11 is expected to come from the M31 halo
population, whereas the contribution by disk stars is expected to be
lower $10 \%$ \citep{bellazzini03}. The metal abundance of G11 has been
calculated both photometrically (see, e.g., \citealt{bonoli87,fan10}) and with integrated spectroscopy   
(see, e.g., \citealt{huchra91,galleti09}) and all results agree
on a moderately low amount of metals. In the following we have adopted 
the most recent spectroscopic re-evaluation of the cluster 
metallicity by \cite{galleti09}: [Fe/H]$_{\rm ZW}=-1.59 \pm 0.23$  and [Fe/H]$_{\rm C09}=-1.60 \pm 0.24$ (on the \citealt{zinn84}, and the \citealt{carretta09}  metallicity scales, respectively).
\citet{rich05} and \citet{federici12} estimated a  reddening   $E(B-V)$=0.10 mag  and 0.12 mag respectively, 
on the line of sight of  G11, and characterized the cluster as having 
a blue HB morphology.

We present in Section 2 the WFPC2 data used in our
analysis. In Section 3 and 4 we discuss the method used to reduce  the
time-series photometry, to detect the variable stars and to derive their periodicities.
In Section 5 we present the RRL sample and the cluster's 
CMD together with the Oo
classification  and the distance modulus of G11. Finally in Section
6 we summarize  our main results. 
\\

%%%%%%%%%%%%%%%%%%%%%%%%% SEC. 2: OBSERVATIONS
\section{Observations}
WFPC2 time-series observations of G11 were obtained  on 2007 July, as part of
the Cycle 15, {\it HST} GO program 11081 (PI: G. Clementini). They consist of 
11 $F606W$ (broad $V$)  and 11 $F814W$ ($I$) exposures each of 1100 s in length,
taken by alternating the two filters. This filter selection allows our data 
to be directly comparable with most previous {\it HST}-based studies of GCs 
and field stellar populations in the MW and in M31, including the M31 field 
RRL stars studied by \cite{brown04,sarajedini09,jeffery11}.
The  exposure length of our observations was the  best compromise to achieve reasonably good signal-to-noise (S/N) ratio even for
measurements corresponding to the minimum light of the M31 RRL 
stars ($V\sim$ 25.5-25.7 mag),  and, at the same time, to avoid smearing the light curves.
These were also the longest possible exposures allowing 
us to fit an $F606W$, $F814W$ pair of images  into one orbit. 
The observations were scheduled in one 11-orbit block with a 
50 minute gap between each pair of observations, for a total exposure 
time on target of 3 hours and 22 minutes in each band, distributed over a total 
time-window of  $\sim 16$ consecutive hours. 
The cluster was roughly centered on the planetary camera (PC) of the 
WFPC2,  which provides better spatial resolution 
(0\farcs 046 pixel$^{-1}$), while
the lower resolution  ($\sim 0\farcs1$ pixel$^{-1}$) 
wide--field cameras sampled the surrounding M31 halo field. 
The WFPC2 has a total field of view (FOV) of 
150$^{\prime\prime} \times 150^{\prime\prime}$, which corresponds to a projected area of $0.6\times0.6$ kpc$^{2}$ at the distance
of M31. 
The tidal radius ($r_{t}$) estimated for G11 is  $\sim 16$ 
arcsec \citep{clementini01}, hence, the region inside the cluster tidal radius is entirely contained  in the PC camera ($34^{\prime\prime} \times 34^{\prime\prime}$).
Our proprietary data were complemented by WFPC2/HST archive observations of G11 
obtained in 2000 as part of the GO program 6671 (PI: Rich). They  
consist of four $F555W$ ($V$) and four $F814W$ ($I$) frames.  The 
individual exposures of the archival data are about 20 minutes long,  and were taken consecutively over 
a total time interval of 5 hours and 15 minutes.
A log of both the proprietary and archival observations of G11 used in the present study is provided in Table~\ref{table0}.

%%%%%%%%%%%%%%%%%%%%%%%%% SEC. 3: DATA REDUCTION
\section {Data Reduction}
Point Spread Function (PSF) photometry of the individual  pre-reduced images of G11 supplied by the STScI pipeline 
was performed with HSTphot (\citealt{dolphin00a}),  a photometry package specifically designed to handle the 
undersampled PSF of the WFPC2/HST data.
A number of initial pre-processing steps are required before running 
the photometry procedure. 
First, image defects such as bad columns and saturated pixels are identified 
using the information provided in the  data quality files. They are then masked by setting them 
 to the bad data value, $-$100 ADU,  and are ignored afterwards.  
Next, cosmic rays and hot pixels are flagged automatically and sky images 
are created in order to provide an initial guess of the background.
At the same time, the X,Y offsets between each image and a reference 
image (for G11 the image u9x6f001m\_c0f in the {\it HST} archive, see http://archive.stsci.edu/)  
 are measured using the IMEXAMINE task in IRAF. These offsets are input 
into the multi-photometry mode of HSTphot in order to measure 
the magnitudes of the objects detected in each image. 
The photometric measurements were performed using PSFs employing the 
``weighted PSF fitting mode'' which places more weight on the central 
part of the star profile and less on the outer pixels.
The photometric procedure automatically applies aperture and Charge Transfer Efficiency 
  corrections
and transforms the instrumental magnitudes to the {\it HST} flight--photometric system 
as well as to the Johnson--Cousins (JC) system, using the transformations available at ${\rm http://purcell.as.arizona.edu/wfpc2\_calib/}$, that 
are based on the calibration formulas by Holtzman et al. (1995). 
In addition to position, magnitudes and errors, HSTphot provides in the output photometry 
file a variety of quality 
information for each detected source,  namely,  the ``object type'' (stellar, extended, bad, etc.), the $\chi^{2}$ 
which simply gives the quality of the PSF fit,  the ``sharpness" used to 
eliminate cosmic rays and extended objects,  the ``roundness",  the S/N ratio, and a 
``crowding'' parameter which describes the change in star brightness 
if neighbors are not subtracted. 
We used the aforementioned quality information to clean the list of detected sources, keeping for the following analysis  only stellar detections 
with valid photometry (object type flag=1) on all input images, global 
sharpness parameter  $-0.3 \leq $ sharpness $\leq$ 0.3, $\chi^2
\leq$ 1.1 if  $V >$ 25.0 mag and  $\chi^2\leq$ 4.0  if $V <$ 25.0 mag, roundness $\leq 2 \times \langle {\rm roundness} \rangle$ and in any case less than 0.45, and 
$ \sigma_{V}$, $\sigma_{I} <0.15$ mag. Fig.~\ref{paramPC} shows the cuts applied to the data of the PC.
The $V, V-I$ CMD   of all sources satisfying these conditions in our 
 Cycle 15 observations of G11 is presented in Fig.~\ref{pippo}.
The limiting magnitudes of our photometry, defined as a 4 $\sigma$ detection, 
are $V \sim$ 27.5 mag in all four cameras, and  $I \sim26.7$ mag and 27 mag on the PC and the WF cameras, respectively. The internal photometric errors in $V$ magnitude and $V-I$ color 
provided by HSTphot are of 
$\sim0.01$ mag down to 
$V \backsimeq 24$ mag; then 
 increase continuously up to $\sim 0.1$ and 0.18 mag, respectively, from 
$V \backsimeq 24$ mag to 27.5 mag. At the level of the HB 
($V\backsimeq 25$ mag) the photometric errors are typically of 
$\sim 0.02$ mag in $V$, and $\sim0.04$ mag in $V-I$.

%%%%%%%%%%%%%%%%%%%%%%%%% SEC. 4: VARIABLE STARS
\section{Identification of the Variable Stars and Period Search}
The identification of the G11 variable stars was based exclusively 
on the Cycle 15 observations. The archive data were used,
when possible, only  for the purpose of improving the period definition and the calibration process.
Identification of the variable stars was carried out using two 
different approaches. First, candidate variables were identified from 
the scatter diagrams of the $F606W$ and $F814W$ datasets, using VARFIND, custom
software developed at the Bologna Observatory by P. Montegriffo.
Briefly, VARFIND computes the standard deviation of the mean magnitude 
of each source over the entire set of individual measurements and compares 
the standard deviation with the  mean magnitude (scatter diagram). Variable 
stars show larger standard deviations 
than constant stars  due to the light variation. 
Fainter stars will also have larger standard deviations due to the photon 
noise, so it is advisable to proceed with caution at low luminosity
levels.

Our second approach was to compute  for each measured source a revised version of the variability index
 (\citealt{welch93}, see also \citealt{stetson96} for a detailed definition and description),  specifically adapted to the HSTphot photometry of the G11 time-series.
Basically, for each star the residual of each individual magnitude measurement with 
respect to the weighted mean magnitude in each photometric band 
is calculated and normalized by the photometric 
error ($R=\frac{x-\bar{x}}{err}$). 
Then, residuals  corresponding to pairs of $V$ and $I$ 
observations closely spaced in time are multiplied 
and finally added ($\sum_{i=1}^n(R_{V_{i}}*R_{I_{i}})$). 
If a star is a genuine variable, it will be brighter than usual in $V$ when it 
is also brighter than usual in $I$, and vice versa. Therefore, the 
total sum of the residual-products will be positive. On the other hand, 
if the star is  constant, and there are only small random errors in 
its magnitudes, then the $V$-band and $I$-band 
residuals will be uncorrelated and should have the same sign half the time, and opposite 
sign the remaining half, leaving the overall sum of the 
residual product close to zero.

The light curves of the candidate variables found with the above procedures were then inspected 
visually.  Period search and study of the light curves were performed using two 
different methods. A first verification of variability and guess of the 
star periodicity was obtained using the phase dispersion minimization
(PDM) algorithm  \citep{stellingwerf78} in the  IRAF environment.  PDM is a 
generalization of the \cite{lafler65} algorithm, and essentially
attempts to identify the phased light curve that produces the minimum scatter
in phase. The period refinement was then obtained with the 
Graphical Analyser of Time Series (GRATIS), a private software developed 
at the Bologna Observatory by P. Montegriffo \citep[see e.g.,][]{clementini00},
which uses both the Lomb periodogram \citep{lomb76} and the best fit of the 
data with a truncated Fourier series \citep{barning63}.
We confirmed the variability, and obtained  
reliable periods and light curves for 14 candidate variables. 
Because of the relatively short-time interval spanned by the Cycle 15 observations our  
periods are generally accurate to two/three significant figures. However, 
six of the confirmed variable stars have a counterpart in the photometric catalogue of the 1999-2000 observations.
For these variables we directly combined the Cycle 15 and the archival $F814W$ datasets.
The long gap between the two  datasets did not cause significant aliasing
difficulties in phasing the $F814W$ data, whereas  there were problems to phase the archive $F555W$  and the $F606W$ data,
but this was mainly  due to the different passbands covered
by these filters. The $F606W$ and  $F555W$ data were thus used separately to iterate the period search procedure. In turn periods 
accurate to five/six significant figures were obtained for the stars with both proprietary and archive time-series (see next Section). 

%%%%%%%%%%%%%%%%%%%%%%%%% SEC. 5: RESULTS
\section{Results}
\subsection{The RR Lyrae stars}
The 14 variable stars identified in the FOV  of the G11 observations are of RR Lyrae type, and include  eight 
fundamental-mode  and six first-overtone pulsators.  The light curves  of the variables in the $F606W$ and  $F814W$ filters  are presented in 
Fig.~\ref{lc}. They are well sampled in both bands. 
The identification, basic elements and photometric properties of the 
confirmed RRL stars in our study are provided in Table~\ref{table2}.
The variable stars are ordered by increasing distance from the center of G11. 
Column 1 indicates the star's ID. Column 2 indicates the  camera of the WFPC2 where the star is located. Columns 3 and 4 provide the star's coordinates in pixels, as measured on the
 reference image (see Section 3).   
Column 5 gives an estimate of the distance
from the cluster center, whereas columns 6, 7 and 8 give the 
RRL type, the star's period, and the Heliocentric Julian date
of maximum light. 
Columns 9  and 10 give the intensity-averaged
mean $F606W$ and $F814W$ magnitude values,
while Columns
11 and 12 list the corresponding $F606W$ and $F814W$ amplitudes of the light
variation, Columns 13 and 14 give the intensity-averaged mean $V$ and $I$ magnitudes, and Columns 
15 and 16 list the corresponding $A_V$ and $A_I$ amplitudes.
\citet{clementini01} identified two candidate RRL stars in G11, based on the 
2000 dataset. 
We confirm the variability
of one of the \citet{clementini01}  candidates that in fact corresponds to the RRc  star V5 in the present study.

The time-series photometry of the variable stars in the $F606W$, $F814W$ HST
 flight-system is provided in Table~\ref{table1},  
which is published in its entirety in the electronic edition of the journal.   
A map showing the position of  the RRL stars detected in the whole 
$\sim2.6\times2.6$ arcmin$^{2}$ FOV covered 
by the WFPC2 observations of G11  is presented in Figure~\ref{fchart}, where we have drawn on the PC a large open circle with the
cluster tidal radius.
RRL stars  falling inside this region  are marked by  filled circles and triangles for RRab and RRc stars,  respectively,
while variable stars outside this region were marked with the corresponding open symbols.  In the figure the 3 circular regions marked as I, II and III on the WF cameras 
enclose the same area contained in the cluster's tidal radius on the PC, and were used to check the contamination by field stars of the cluster CMD (see Section 5.2). 
Fig.~\ref{pc_fchart} provides the finding chart of the seven variables detected on the PC.

In order to plot the variable stars in the CMD and in the period-amplitude diagram 
  we have converted the {\it HST} 
flight-system magnitudes into the Johnson-Cousins photometric system.
The calibration  of the $F606W$, $F814W$ light curves  to the Johnson-Cousins $V,I$ was performed through a procedure 
that properly takes into account the variation in color of the variable stars during the
pulsation cycle. This procedure, which we developed and widely tested in 
 a number of other studies (see, e.g., \citealt{baldacci05,fiorentino10a} for previous applications with ground-based and {\it HST} data, respectively)  basically performs the 
following steps:  {\it i)} to each single phase--point in a given filter associates the color at
the corresponding proper phase read from the Fourier models best fitting the star light curves\footnote{Depending on the sampling of the light curves, from 1 up to 3 harmonics 
were used to fit the RRc light curves, whereas from 2 up to 5 harmonics were used the RRab stars.} obtained 
with the GRATIS package; {\it ii)} this color is then entered into the calibration
equations (in the present case \citealt{holtzman95} equations) thus obtaining individual $V, I$ calibrated data.
The $V,I$-calibrated light curves are then re-analyzed with GRATIS (by imposing the final period, epoch of maximum light and optimal number of harmonics 
defined during the period search process) in order to compute the intensity-averaged mean magnitudes and the amplitudes listed in Table~\ref{table2}.
We explicitly note that, although HSTphot provides mean magnitudes of all photometrized stars in the 
Johnson--Cousins system, these values are  not appropriate for variable stars for two reasons:
1)  for variable stars the mean
magnitude strictly depends on the sampling of the light 
curve, making it  necessary to use models/templates in case of poor sampling; 2) HSTphot provides mean magnitudes computed as the average of the magnitude values,
 however, intensity-averaged mean magnitudes are needed for  the variable stars,  since this is the kind of  mean value that better approximates 
 the luminosity  the variable star would have were it  non variable (see, e.g.,  Caputo et al. 2000).
The 14 confirmed variable stars are plotted in the CMD, using their intensity-averaged mean magnitudes and colors,  and filled/open circles and triangles for RRab and RRc stars within/outside the cluster tidal radius, respectively.
According to Fig.~\ref{fchart} seven of the 14 RRL stars fall inside the PC FOV, five being 
within the cluster  tidal radius 
 and two (V6 and V7) located slightly outside $r_{t}$. The position in the CMD of these two variables
suggests that V6 is likely a cluster member, while V7 lies slightly below the cluster HB  and could possibly be a background interloper. To be conservative, in the following we will 
consider also V6 as a field star, however we note that including the star in 
the cluster RRL sample would not affect the results of our analysis.
The remaining seven RRL stars are placed in more external regions and may in part belong 
to the M31 halo (see Sect.~\ref{ss:oo}).
Although the number of RRL stars found within the cluster tidal radius is fairly small, this result is 
consistent with the rather blue HB morphology exhibited by G11 
\citep[see Figure~11 in][recall that predominantly blue HB clusters often contain few, if any, 
RRL variables]{rich05}, and it is also consistent  with the \cite{clementini01} 
results. 

\subsection{Color--Magnitude Diagram}\label{ss:cmd}

In Fig.~\ref{cmds_f}  we have divided the stars in the CMD  of Fig.~\ref{pippo} 
in three regions at increasing distance from the 
center of G11, in order to better  isolate the cluster stars from the M31 field population.
%left panel
The left panel of the figure corresponds to the region within the tidal radius of G11.
The CMD of this region shows two evolutionary sequences that clearly trace the cluster stars:
{\it i)} a rather steep red giant branch (RGB), indicating
that G11 is relatively metal poor, in agreement
with the metallicity values found in the literature \citep{rich05,galleti09,federici12};
and {\it ii)} an extended blue HB, which is absent in the 
control field shown in the central panel of  Fig.~\ref{cmds_f}.
To check whether these extremely blue stars do indeed belong to the G11 HB, in Fig.~\ref{cmds} 
we compare the CMD of the stars within the cluster tidal radius on the PC with the CMDs 
of stars located in the three equal-area external regions of the WF cameras shown in  
Fig.~\ref{fchart}. 
The number of extremely blue stars is significantly larger in the region within the G11 
tidal radius (9) than in field I (2), II (1), and  III (1; see  Fig.~\ref{cmds}),
thus confirming that these stars likely belong to the G11's HB. 
These two features, along with the mere presence of RRL stars, lead to conclude that 
G11  is older than 10 Gyr and most likely of Oo~II type, as confirmed by the mean 
period of its RRab variables (see Sect. \ref{ss:oo}). 

%mid panel
The middle panel of Fig.~\ref{cmds_f} shows an intermediate region 
of our FOV, whose limits were chosen in order to enclose 
the same area covered by the G11 tidal radius and thus easily estimate the field 
contribution to the cluster CMD.
The field contamination does not appear to be a significant issue,  
being negligible at the HB level. No RRL stars  have been found in this region. 

%right panel 
Finally, in the right panel of the figure, we show the outermost 
region %($r>40^{\prime\prime}$) 
of our FOV, which better traces the M31 
halo stellar population. In fact, we can clearly see some well--known 
features of M31: a broad RGB with a descending tip, and an extended red clump 
located at $V \sim 25.0-25.5$ mag and $V-I \sim 1$ mag. 
We can also easily distinguish a minor sign of blue stars (with color
$\le 0.5$ mag) which may be the contribution of main sequence
intermediate--age stars (blue plume) possibly associated with the
galaxy disk, and/or  blue HB stars belonging to the old field population
associated with the galaxy's halo.
%%%%%%%%
An old population (older than 10 Gyr) is confirmed by the 
seven RRL stars we have identified in this region. 
Figure~\ref{cmdhb}  shows an expanded view of the HB: field and cluster 
variables fall reasonably well in the appropriate region of the HB according to
their pulsation modes  and  colors.  
Only exceptions are the RRab star V4 and the RRc star V5 among the cluster  
variables, which  are respectively slightly too blue and  too red for their 
pulsation modes, and the field RRab star V11 which is slightly too blue.  
We note also that V4 and  V1 are  about 0.1 - 0.2 mag 
brighter than all the other cluster variables. We suspect that they might be 
contaminated by unresolved companions.
On the other hand, the field RRab star V7 is slightly fainter than all the other 
variables, suggesting that it might be a background object. 
Fig.~\ref{cmdhb}  also shows that  a number of objects  
lying inside  the IS region  do not show any RRL-like variation. 
We believe that these objects 
are the result of
blending in the dense core region, mainly between blue HB stars and red
giants \citep[see][for a detail discussion of this effect]{rich05}. 
Another possibility is also that they are back/foreground 
field stars.\\                              

%At these large distances (almost three times the $r_t$ of G11), 
%these RRLs  may in part belong to the field halo population of M31 (see discussion 
%at the end of Sect. \ref{ss:oo}).   }

\subsection{The Oosterhoff type of G11}\label{ss:oo}

The average period of the three RRab stars located within the G11 tidal radius  is 
$\langle P_{ab} \rangle =0.70$~d ($\sigma =$0.09~d, average on 3 stars), suggesting 
a classification of G11 as Oo type II.
Similarly, the average period of the two RRc stars: $\langle P_{c} \rangle=0.40$ d 
($\sigma =$0.04~d, average on 2 stars), and the fraction of {\it c-}type RR Lyrae, 
$f_{c}=0.4$, are also consistent with an 
Oo~II classification (see Section \ref{intro}). 
However, \citet{catelan12} note that neither 
$\langle P_{c} \rangle$ nor $f_{c}$ appear to be good Oosterhoff
indicators  and suggest that $P_{ab,min}$, besides $\langle P_{ab}
\rangle$, is a quantity that better defines 
the Oosterhoff status of a stellar system. In the case of G11, the 
shortest-period RRab (within the cluster tidal radius) is star V3 with
$P_{ab,min} = 0.626$~d, which also suggests 
an Oo~II classification, albeit within the limits of poor statistical significance.
 
%%%%%%%%%%%%%%%%%%%%%%%%% Rab vs Met
 In Figure~\ref{oo}, we show  the run of  $\langle P_{ab} \rangle$ values vs metallicity 
for  the  GGCs containing RRL stars, using for the clusters' metallicity the \citet{zinn84} 
scale in the left panel, and the \citet{carretta09} scale in the right panel.
 The figures illustrate the well--known dichotomy presented by the  MW GCs:
moderately metal--poor GCs tend to  group around $\langle P_{ab} \rangle \sim
0.55$ d, while very metal--poor GCs  preferentially
have $\langle P_{ab} \rangle \sim 0.65$ d. Clearly visible is  the 
region avoided by the GGCs, which spans the period range   
$0.58 \lesssim \langle P_{ab} \rangle \lesssim 0.62$ d and is 
usually referred to as the ``Oosterhoff gap'' \citep[e.g.][]{catelan04,catelan09}. 
%Also shown in the figure are the so-called Oo~III clusters, namely,  NGC~6388 and NGC~6441. 
%In spite of the high metal abundance
%These two GCs have the red HB characteristic of metal-rich GCs but also quite extended blue HBs 
% \citep{busso07},  and contain fundamental-mode RRLs with extraordinary long periods (\citealt{corwin06}, and references therein), 
%thus setting aside from the other MW GCs in the  $\langle P_{ab} \rangle$ {\it vs} [Fe/H] plane. 
  According to  the metallicity adopted in the present study ([Fe/H]=$-$1.59 or $-$1.60 dex 
in the \citealt{zinn84} and the \citealt{carretta09} metallicity scale, respectively),  
G11 (asterisk in Fig.~\ref{oo}) locates in 
the ``transition'' region where, in the MW, one finds GCs on both 
sides of the Oosterhoff-gap, i.e, GCs which have almost the same
metal abundance, yet very different $\langle P_{ab} \rangle$ values.
The rather long  mean period of its RRab stars clearly places G11 in the Oo~II 
region of the plot, in perfect agreement with its blue HB.

%%%%%%%%%%%%%%%%%%%%%%%%% BAILEY
The Bailey (period--amplitude) diagram has also been proposed as an indicator of 
the Oosterhoff typology \citep{clement99}. 
In this diagram, RRL stars belonging to the two 
different Oosterhoff groups appear well separated: for a given amplitude (which
corresponds to a given temperature) Oo~II RRL stars  show longer periods than Oo~I variables.
The usually adopted Oo~I locus has been derived using RRL stars belonging
to the GC M3 and both the linear (Clement \& Rowe 2000) and the non-linear
\citep{cacciari05} relations well represent most of the Oo~I GGCs. 
The Oo~II relation is less well defined. 
\cite{clement00b} derived an Oo~II locus using the metal-poor  RRL stars in the 
GC $\omega$ Centauri. However, this cluster is well known to be a peculiar object 
(see e.g. \citealt{sollima05}) which  might be   
the core of a stripped dwarf spheroidal galaxy 
(e.g., \citealt{dinescu02,lee02,altmann05,meza05,bekki06,villanova07,wylie10,majewski12}). 
On the other hand, \citet{cacciari05} found that their Oo~I quadratic
relation, shifted by $\Delta$ log P $\sim$ 0.06
d (at fixed amplitude) 
to match the evolved stars in M3, also
corresponds approximately to the mean locus traditionally assigned to the Oo~II
variables \citep{sandage81b}. However, we suspect that none of these
relations is fully representative of the Oo~II GGCs.  
In order to compare the M31 GCs with a relation that better represents the Oo~II GGCs, 
Contreras Ramos (2012, in preparation) obtained a new Oo~II locus by fitting the
periods and amplitudes of more than 190 fundamental-mode RRL stars 
belonging to the 19 Oo~II clusters shown in Fig. \ref{oo} (right panel),  
excluding known Blazhko (\citealt{blazhko1907}) and blended stars. 
The new Oo~II locus is represented by the linear
relation: $A_{V}{\rm (RRab)} =0.15( \pm 0.05) - 4.10 ( \pm 0.25) \log P_{ab}$ 
(with  $\sigma = 0.16$ mag and a correlation coefficient of 0.76)  and is plotted as a 
solid line in the left panel of Figure~\ref{bai} along
with the RRL stars detected in the present paper. 
%and  a grey-shaded region corresponding to  its 1-$\sigma$ confidence levels. 
%The full analysis including non linear relations will be presented in a
%forthcoming paper (Contreras Ramos 2012, in preparation).  
%Also shown in the figure are the loci of the Oo~I and Oo~II GGCs according to 
%\cite{clement00b} and the quadratic Oo~I locus  by \cite{cacciari05}. 
The position of the G11 RRab stars is very consistent with the new Oo~II line.
In the right panel of Fig.~\ref{bai} we compare the G11 RRL stars  with the individual
variables belonging to the Oo~II GGCs with most similar metallicity, as shown 
in Fig. \ref{oo} . 
%sharing the same position of G11 in Fig.~\ref{oo}, namely, NGC~5986, M2, M9, NGC~4388, M55, and M79. 
%(M22 was omitted because only photographic data are available for its variable stars).   
Fundamental-mode and first overtone RRLs in all these clusters occupy the same regions 
of the Bailey diagram with only very few exceptions.
% once again confirming the Oo~II nature of G11 and the cluster similarity to its MW counterparts. 
In conclusion, according to all the pulsation indicators ($<P_{ab}>$, $<P_{c}>$, 
$N_{c}/N_{ab}$, and location in the $<P_{ab}>$-[Fe/H] and period-amplitude planes) 
G11 appears to be a pure Oosterhoff type II GC, comparable with 
the GCs with similar properties (HB morphology and [Fe/H]) observed in the MW. 

Three of the variable stars outside the tidal radius of G11 fall close to the Oo~I locus 
(see left panel of Fig.~\ref{bai}).  They are: V11, V12 and V13. 
V11 and V12 have short periods of 0.51 and 0.43 days, respectively, which are consistent 
with an Oo~I classification. 
They are among the most external RRLs we have identified in our FOV and likely belong 
to the field of M31, which is know to have Oo~I properties 
(see, e.g., \citealt{brown04,sarajedini09, jeffery11}).  
%
%No direct estimate of the number of  RRLs we can expect in the M31 field around G11 exists, as previous studies 
%have either monitored regions of M31 closer to the M31 disk (\citealt{brown04,sarajedini09}), or the M31 halo on the opposite side with respect
%to G11's 
%location (see \citealt{jeffery11} and Fig.~1 of  \citealt{clementini11}). 
%However, if we assume that Jeffery et al.'s halo fields (fields H21 and H35b, respectively)  provide a good representation
%of the M31 halo in general  and normalize their RRL number density to the WFPC2  area outside the G11 tidal radius we can expect to find 1-2 field RRLs  around G11, 
%thus supporting the hypothesis that V11 and V12 may indeed be field  stars. 
%
The period of V13, $P$= 0.65 days, suggests an Oo~II classification, but the rather 
small visual amplitude places this star closer to the Oo~I line. 
However, the visual amplitude of V13 appears to be smaller than its red $I$-band amplitude. 
%suggesting that the star may be contaminated by a companion. 
In Fig.~\ref{baii} we show the $I$-band period-amplitude diagram of all the G11 RRL stars 
and compare it with the Oo~II locus by \citet{arellanoferro11}.  
The position of V13 appears to be fully consistent with the $I$-band Oo~II locus, 
thus suggesting that the star may be contaminated by a companion and its 
$V$-band amplitude reduced by blending. 
The same conclusion can be applied to  V4, which was noted in Sect. \ref{ss:cmd} to be too blue 
for its period and brighter 
than the other G11 RRL stars. %, also seems to indicate that some blending  
%might be affecting the $I$ amplitude of the star. 

\subsection{Distance to G11}\label{ss:dist} 

The distance to G11 was derived, along with the cluster metallicity and reddening,  
by fitting the cluster red giant and horizontal 
branches  to the ridgelines of well--studied GGCs. 
Fig.~\ref{fit_griglia} shows the results obtained by fitting  the CMD of G11, from 
the sources imaged on the PC,  to the ridgelines of the GGCs 
47 Tuc ([Fe/H]=$-$0.71/$-$0.76 dex, where the first metallicity value is on 
the \citealt{zinn84} scale and the second value is on the \citealt{carretta09} scale);  
M5  ([Fe/H]=$-$1.4/$-$1.33 dex);  M3 ([Fe/H]=$-$1.66/$-$1.50 dex); 
M15 ([Fe/H]=$-$2.15/$-$2.33 dex); and 
M92 ([Fe/H]=$-$2.24/$-$2.35 dex).  
References for the 
GGC CMDs can be found in \citet{federici12} who also provide an exhaustive description  
of the fitting procedure. 
% that they 
%applied to the 2000 sub-set of G11 data analyzed in this paper obtaining a very similar 
%result to ours. 
The  CMD of G11 is best fitted by the ridgeline of the GGC M3 for $E(B-V)$ = 0.12 mag, 
 and a distance modulus of  
$(m-M)_0 = 24.46 \pm$ 0.10 mag (corresponding to $D$ = 780 $\pm$ 72 kpc), in good 
agreement, within the errors, with the M31 modulus derived by   
\cite{riess12}  and the distance to G11 derived by \cite{federici12}.
Fig.~\ref{fit_griglia} shows that the average luminosity of the G11 RR Lyrae stars on 
the PC is reasonably well fitted by the M3 HB, however, as 
pointed out at the end of Sect. \ref{ss:cmd}, two of the cluster RR Lyrae stars (namely, 
V1 and V4) are rather overluminous, and the remaining three 
variables within the cluster tidal radius (V2, V3 and V5) also lead to an average 
luminosity   of $\langle V(RR) \rangle$ = 25.20 mag ($\sigma$=0.07 mag, average on 3 
stars) that is about 0.17 mag brighter that the mean HB magnitude obtained by fitting 
the cluster CMD with the ridgeline of M3. 
Blending and contamination by companions in the crowded central regions of G11 may 
be responsible for the high luminosities of  the RR Lyrae stars within the cluster 
tidal radius, particularly in the 
case of  V1 and V4. In this respect it is  worth of notice that $\langle V(RR) \rangle$ 
becomes fainter with increasing distance from the cluster center (see Fig.~\ref{vhb}), 
and in fact the
average magnitude of the 4 most external variables (namely, V11, V12, V13 and V14) 
is  25.38 mag ($\sigma$ = 0.05 mag, average on 4 stars), in good agreement with the 
value of $\langle V(HB) \rangle$ = 25.37 mag obtained by the fit with M3. 
Furthermore, the rather blue HB morphology of G11 also suggests that the cluster  
RR Lyrae stars might be  overluminous because they are evolved off the Zero Age 
Horizontal Branch (ZAHB). 
Both the aforementioned effects  may conspire to produce the enhanced luminosities 
observed for the G11 RR Lyrae stars and caution against using these variables 
to estimate the cluster distance.

%%%%%%%%%%%%%%%%%%%%%%%%% SEC. 6: CONCLUSIONS
\section{Summary and Conclusions}
We have performed a study of the variable star population of the globular cluster G11 based on WFPC2/HST time-series 
data properly scheduled  in order to detect and characterize the RRL
variable star population of this M31 cluster. A main purpose of this study was to verify whether the Oosterhoff  
dichotomy exists in the Andromeda galaxy. 
We summarize our results as follows:
\begin{itemize}   
\item We have identified 14 variable stars, all belonging to the RRL
  class,  in the 150$^{\prime\prime} \times 150^{\prime\prime}$ total FOV covered by our WFPC2 observations. 
  Five of them (3 RRab's and 2 RRc's) are placed within the tidal radius of G11 (e.g., $r < 16\arcsec$) thus making them very likely
  cluster members. The remaining RRL  stars are located at distances in the range of 17\farcs7 to 
  98\farcs7 from the cluster center, and may in part represent  the M31  field variable star population; 

\item The G11 CMD shows  that the cluster HB  is quite extended in color, stretching across the IS,
and extending significantly to the blue through a tail that reaches 
$V \sim 26$ mag at $V-I  \sim 0.2$ mag, shows a gap, and then an
extremely blue tail formed by 
stars with V $\geq 26.3$ mag and V-I $\lesssim 0.2$ mag,  likely belonging to the G11's HB,
since this same feature is lacking in the control fields;

\item The distance to G11 was derived 
  by fitting the cluster red giant and horizontal branches,
 from the sources imaged on the PC,  to the ridgelines of well--studied GGCs. The  CMD of G11 is best fitted by the ridgeline of the GGC M3 for 
 $E(B-V)$ = 0.12 mag, 
   and a distance modulus of  
$(m-M)_0 = 24.46 \pm$ 0.10 mag. 

\item The average luminosity of the G11 RRL stars on the PC is reasonably well fitted by the M3 HB; however, two of the cluster RRL stars (namely, V1 and V4) 
are rather overluminous, and the remaining three variables within the cluster tidal radius also lead to an average luminosity  that is about 0.17 mag brighter that the mean HB magnitude obtained by fitting the cluster CMD with the ridgeline of M3.
Blending and contamination by companions in the crowded central regions of G11, along  with evolution off a position on the blue ZAHB, 
given the rather blue HB morphology of G11,
likely conspire to produce the enhanced luminosities observed for the G11 RR Lyrae stars and caution against using these variables 
to estimate the cluster distance. 
\item Using the three  RRab cluster members, we find $\langle P_{ab} \rangle =0.70$~d 
  and $P_{ab,min} = 0.626$~d. Both quantities qualify G11 as an Oo~II GC. The position of G11 in the Period-[Fe/H]  and period-amplitude diagrams
  further supports its Oo~II nature and confirms the importance of 
  the HB morphology in the Oosterhoff classification.
\end{itemize}   

To conclude, we have found the first evidence that Oosterhoff type II GCs 
exist in the Andromeda galaxy and seem to follow the same rule of their MW counterparts, where
moderately metal--poor ([Fe/H] $\sim -1.6$) GCs with a  blue HB, and containing RRL
stars show Oo~II properties.

\bigskip
\acknowledgments 
We thank the referee, Prof G. Preston  for his comments and suggestions that helped to improve the manuscript.
We are indebted to the Program Coordinator, A. Roman, and the Contact Scientist, M. Sirianni, of our {\it HST} program for their invaluable help with the Phase II and scheduling of the {\it HST} observations. A special thanks goes to P. Montegriffo for the development
and maintenance of the VARFIND and GRATIS softwares and G. Greco for useful discussion on the
statistical significance of the new Oo~II linear correlation. GF has been supported by the  INAF fellowship 2009 grant.
Financial
support for this research was provided by COFIS ASI-INAF I/016/07/0, and 
by the agreement ASI-INAF I/009/10/0.  HAS and BJP thank the Space Telescope Science
institute for support under grant HST-GO-11081.05-A. HAS also thanks the NSF for
support under grant AST 0707756.  Support for MC is provided by the Chilean Ministry for the Economy, Development, and Tourism's Programa 
Iniciativa Cient\'{i}fica Milenio through grant P07-021-F, awarded to The Milky Way Millennium Nucleus; by Proyecto Fondecyt Regular \#1110326; by the BASAL Center for Astrophysics and Associated Technologies 
(PFB-06); and by Proyecto Anillo ACT-86. RCR acknowledges partial support by Proyecto Fondecyt Postdoctorado \#3130320.

\bibliographystyle{apj}
%\bibliography{biblio}

\newpage

%%%%%%%%%%%%%%%%%%%%%%%%%%  FIGURES %%%%%%%%%%%%%%%%%%%%%%%%%
 
%%%%%%%%%%%%%%%%%%%%%%%%%% Fig1: GC Targets
\begin{figure*} [b]
\centering
\includegraphics[width=16.3cm,clip]{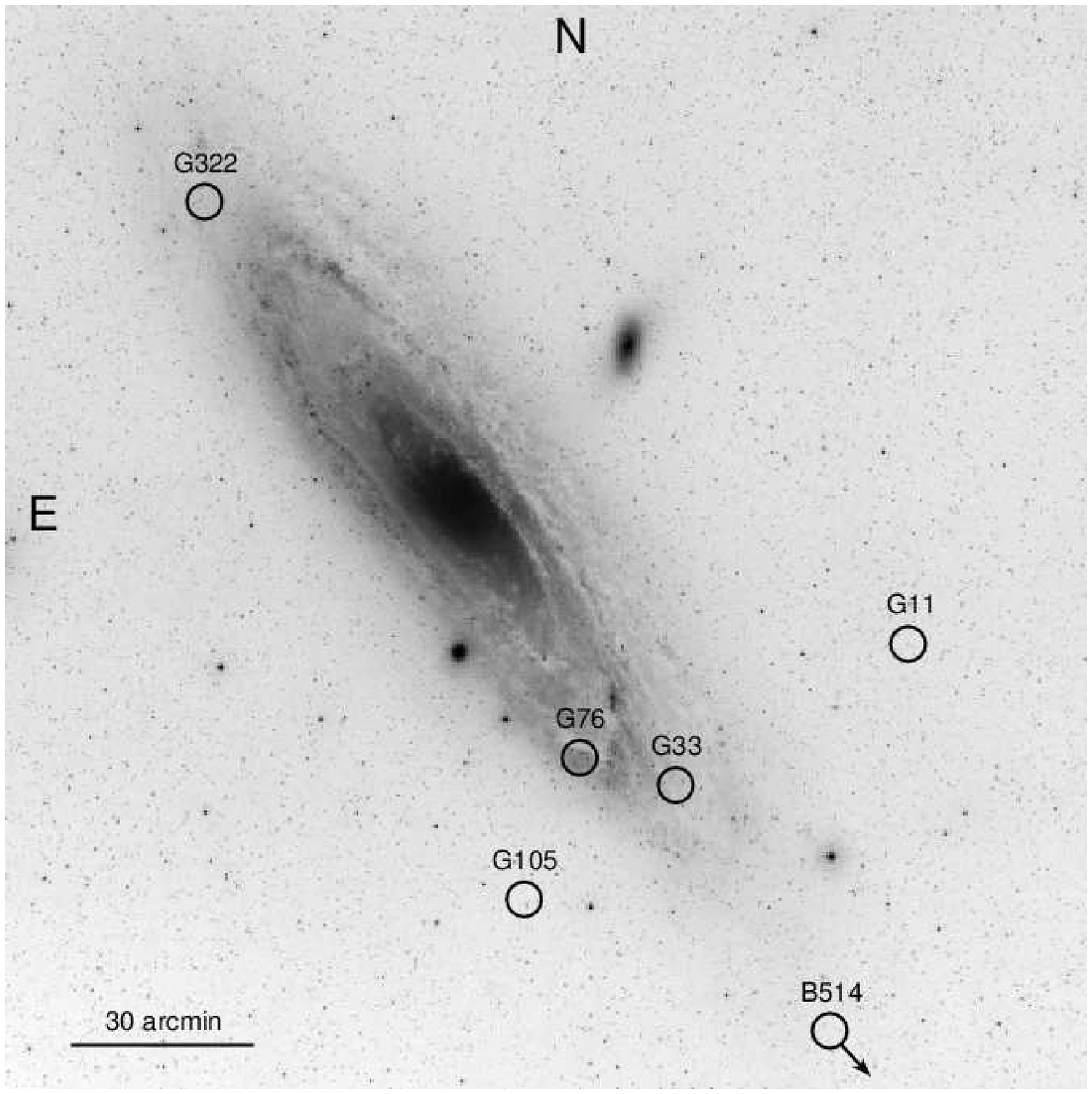}
\caption{Location of the GCs targeted in the Cycle 15 {\it HST} program GO 11081 on a  $\sim 2.9 \times 2.9$ deg$^2$  POSSI/O/DSS2
 image of the Andromeda galaxy from the Digitized Sky Survey  available at CDS (Strasbourg).
Clusters G11, G105 and B514 are located in the M31 halo, whereas G76, 
G33 and G322 are projected on the disk component of M31. B514 is the
farthest  GC in our sample at a distance of  $\sim4$ degrees from the center of M31, and its 
real position does not appear in this map.
}
\label{targets}
\end{figure*}

%\clearpage

%%%%%%%%%%%%%%%%%%%%%%%%%% Fig2: paramPC
\begin{figure*} 
\centering
\includegraphics[width=16.3cm,clip]{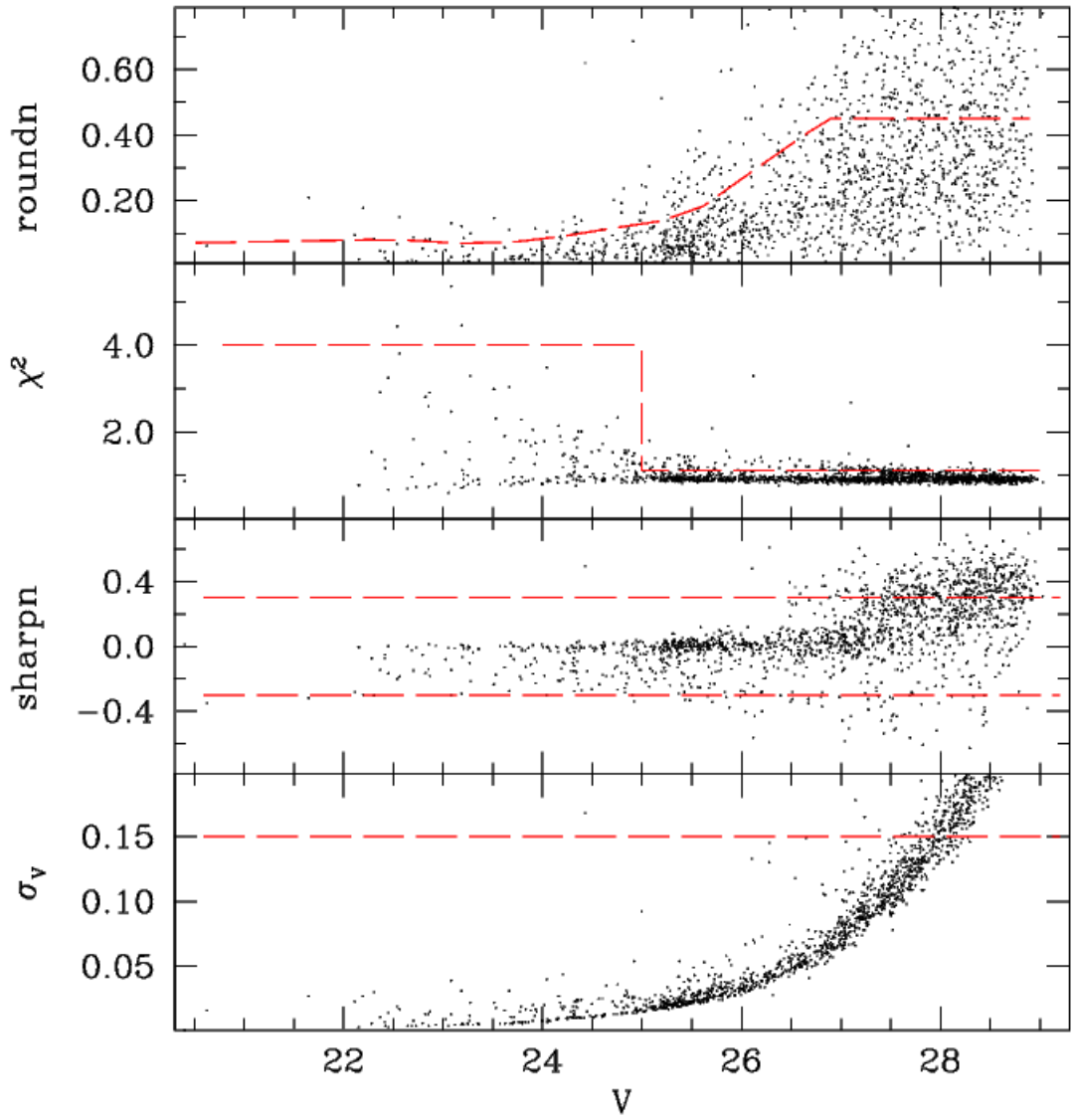} 
\caption{HSTphot quality image parameters of the G11 data  for the  PC. The dashed lines show the cuts we have applied to the photometric
catalogues.}
\label{paramPC}
\end{figure*}

%\clearpage

%%%%%%%%%%%%%%%%%%%%%%%%%% Fig3: CMD totale
\begin{figure*} 
\centering
\includegraphics[width=16.3cm,clip]{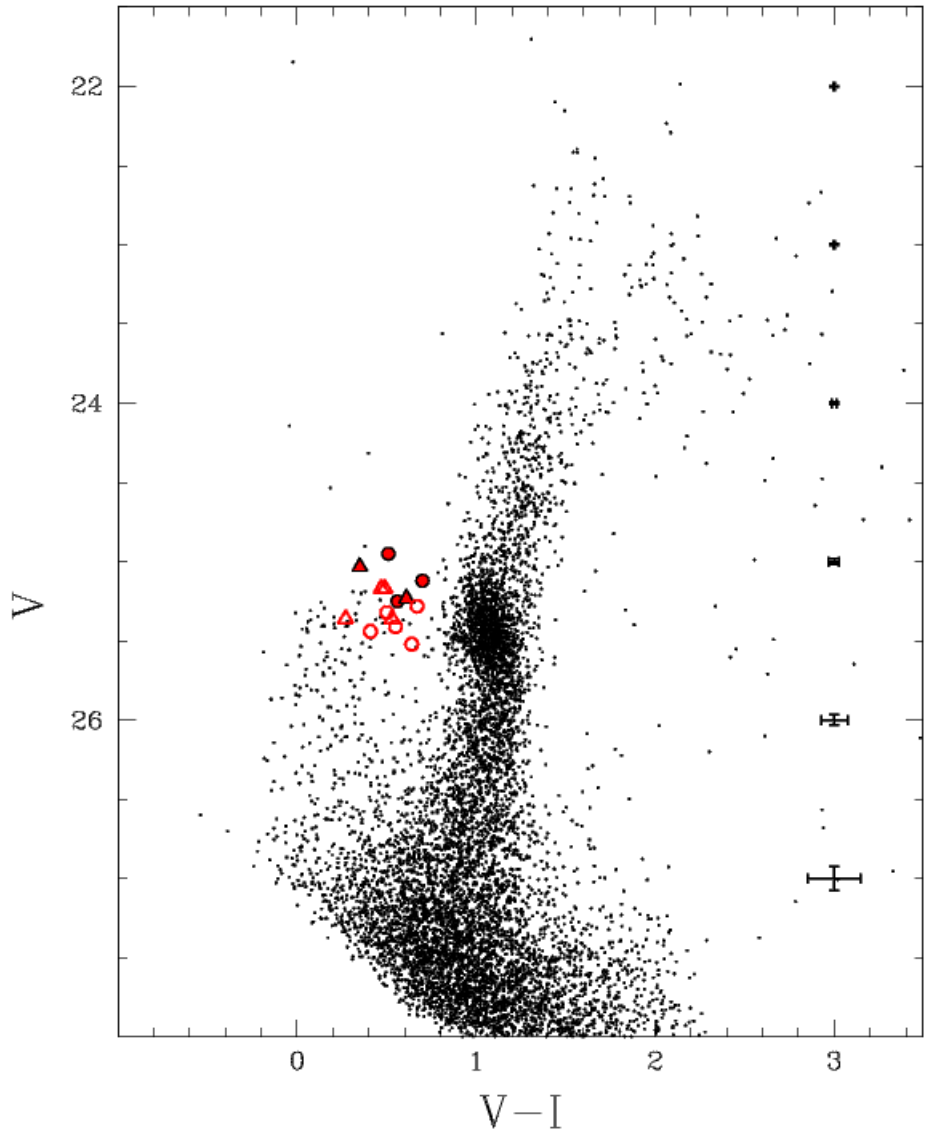} 
\caption{$V, V-I$ CMD of the stellar sources  
(object type flag=1)  in the FOV of
the Cycle 15 observations of G11, selected as described in Sect.~3. 
 Circles and triangles mark respectively fundamental-mode  (RRab) and first-overtone (RRc) RRL stars  we have identified in the whole FOV
 of the G11 observations, 
with filled and open symbols corresponding to variable stars within and outside the cluster tidal radius, respectively (see text for details). 
}
\label{pippo}
\end{figure*}

%\clearpage

%%%%%%%%%%%%%%%%%%%%%%%%%% Fig5a-b: LIGHTCURVES 
\begin{figure*} 
\centering
\includegraphics[width=16.3cm,clip]{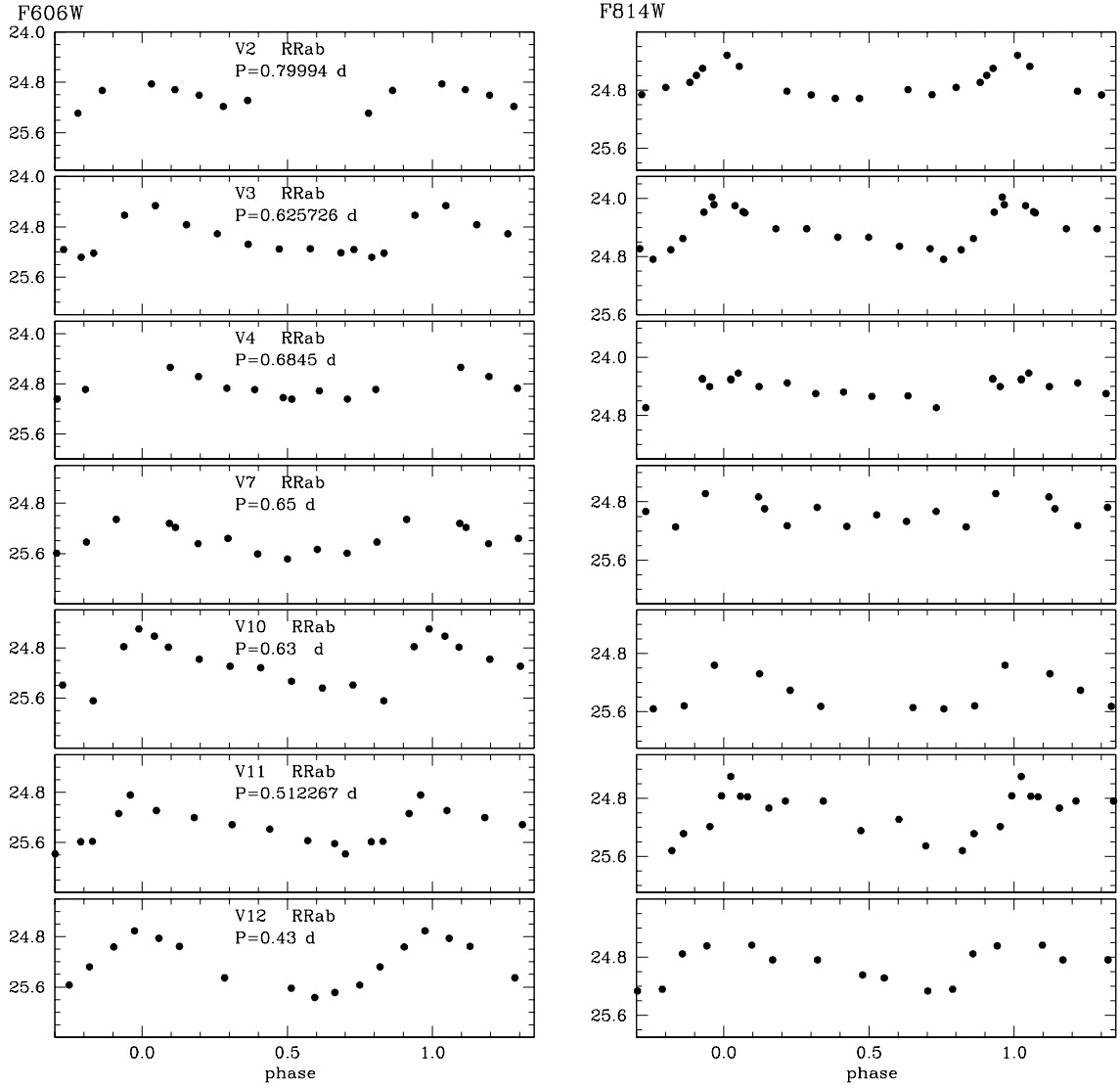} 
\caption{$F606W$, $F814W$ light curves of the  RRL stars detected in this study, 
ordered by increasing distance from the cluster center (R.A. = {\rm 00$^h$ 36$^m$ 20\fs86;  
decl.(J2000) = 40$^{\circ}$ 53$^{\prime}$ 37\farcs25}). Typical errors of the single data points are of about 
0.08 mag and 0.13 mag in $F606W$ and $F814W$, respectively.}
\label{lc}
\end{figure*}

\clearpage

\begin{figure*} 
\centering
\figurenum{4}
\includegraphics[width=16.3cm,clip]{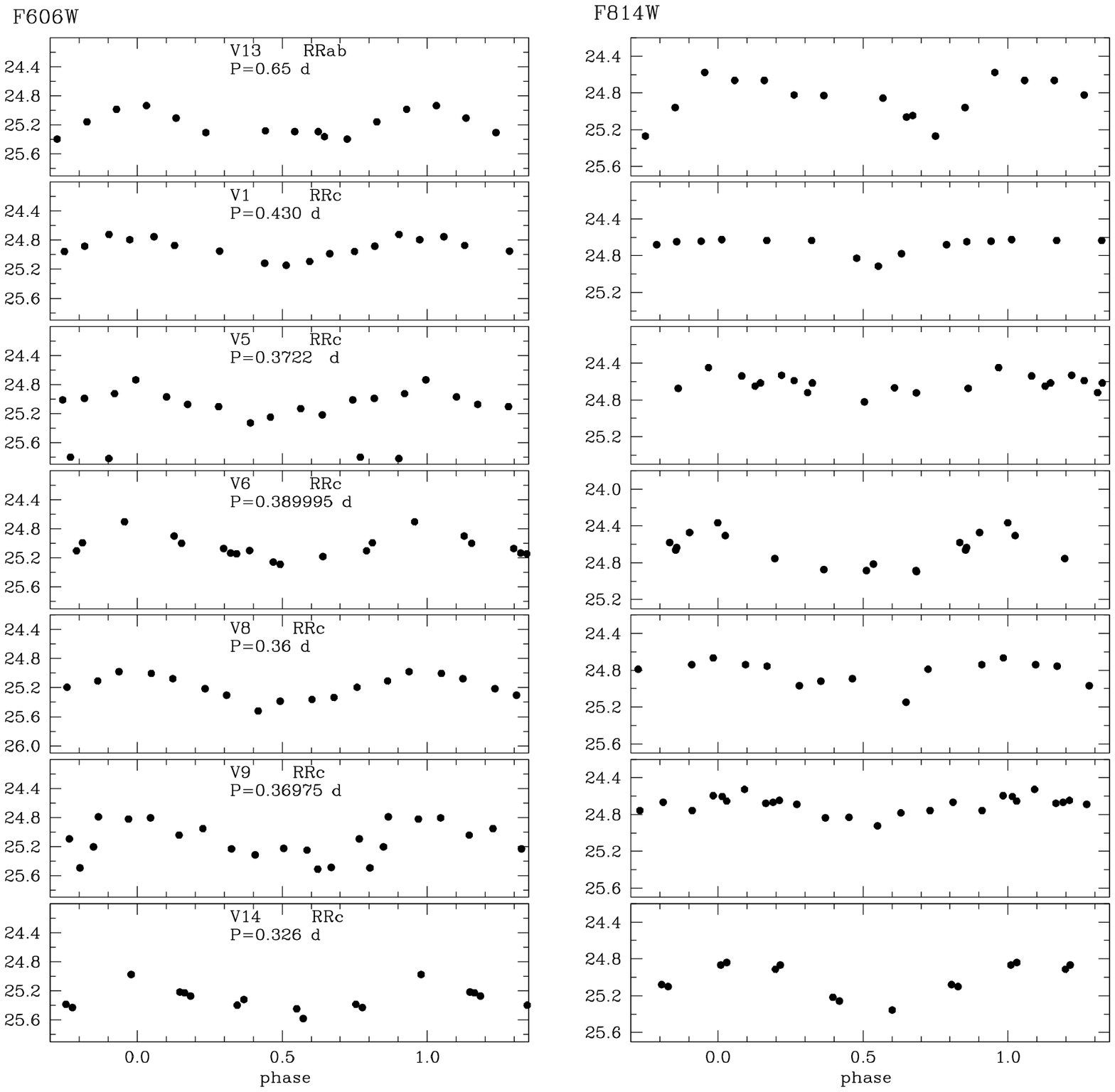} 
\caption{ continued --}
%\label{lc_c}
\end{figure*}

\clearpage

%%%%%%%%%%%%%%%%%%%%%%%%%% Fig6: FOV
\begin{figure*}
\centering
\includegraphics[width=16.3cm,clip]{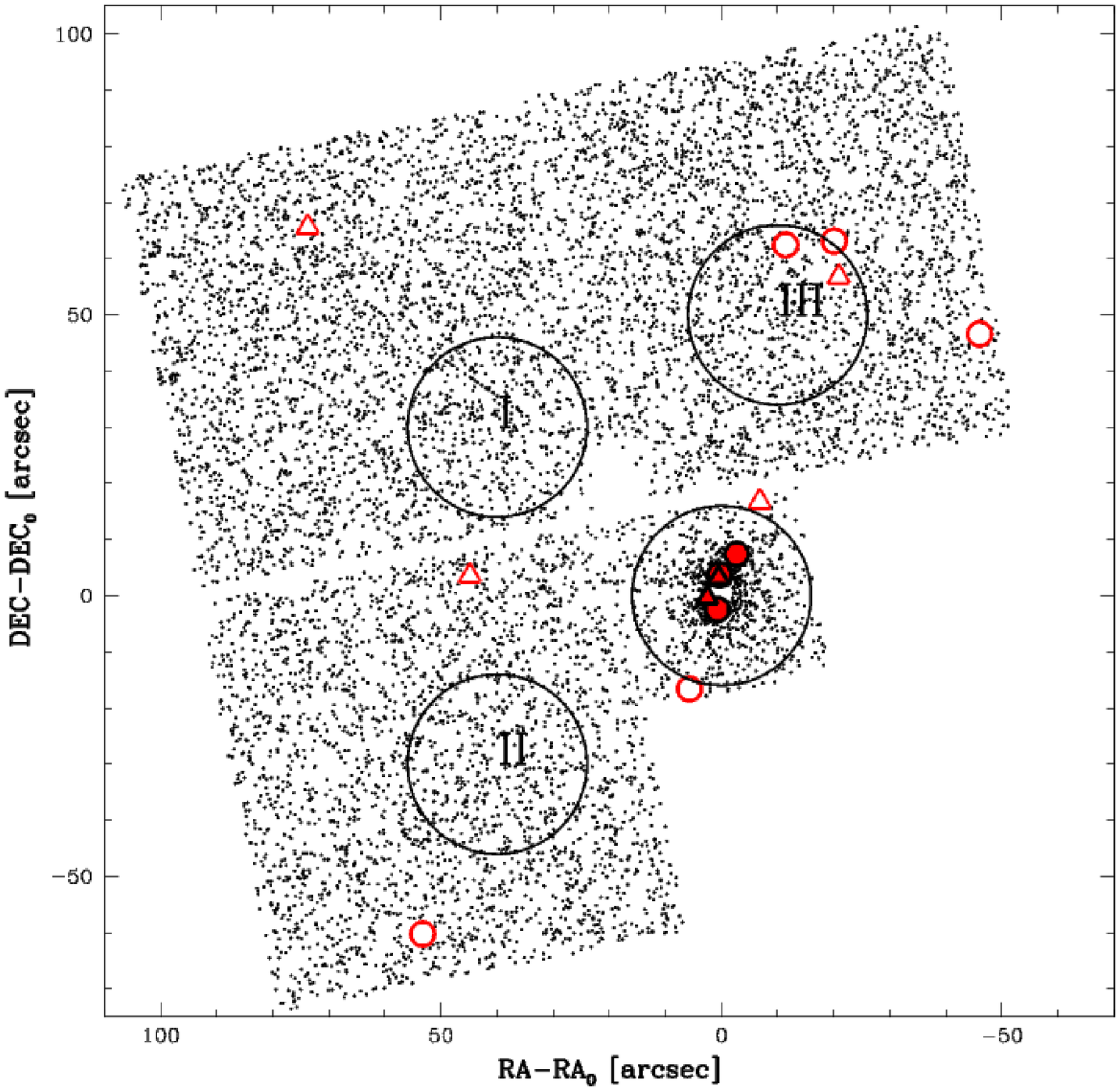} 
\caption{Map of the stellar sources  
 in the  $\sim$ 2.6$^{\prime} \times$ 2.6$^{\prime}$  FOV covered by the WFPC2 observations of G11.  
 Black points mark non-variable sources. RRab and RRc stars are marked by  
 circles and triangles,  respectively. Filled (red) symbols
indicate the likely cluster members, whereas empty symbols indicate RRL stars  
outside the cluster tidal radius. 
The large empty circles on the 4 cameras of the WFPC2 are  drawn with the tidal
radius of G11 as derived by \cite{clementini01}.  
}
\label{fchart}
\end{figure*}

\clearpage

%%%%%%%%%%%%%%%%%%%%%%%%%% Fig7: PC_FCHART
\begin{figure*}
\centering
\includegraphics[width=16.3cm,clip]{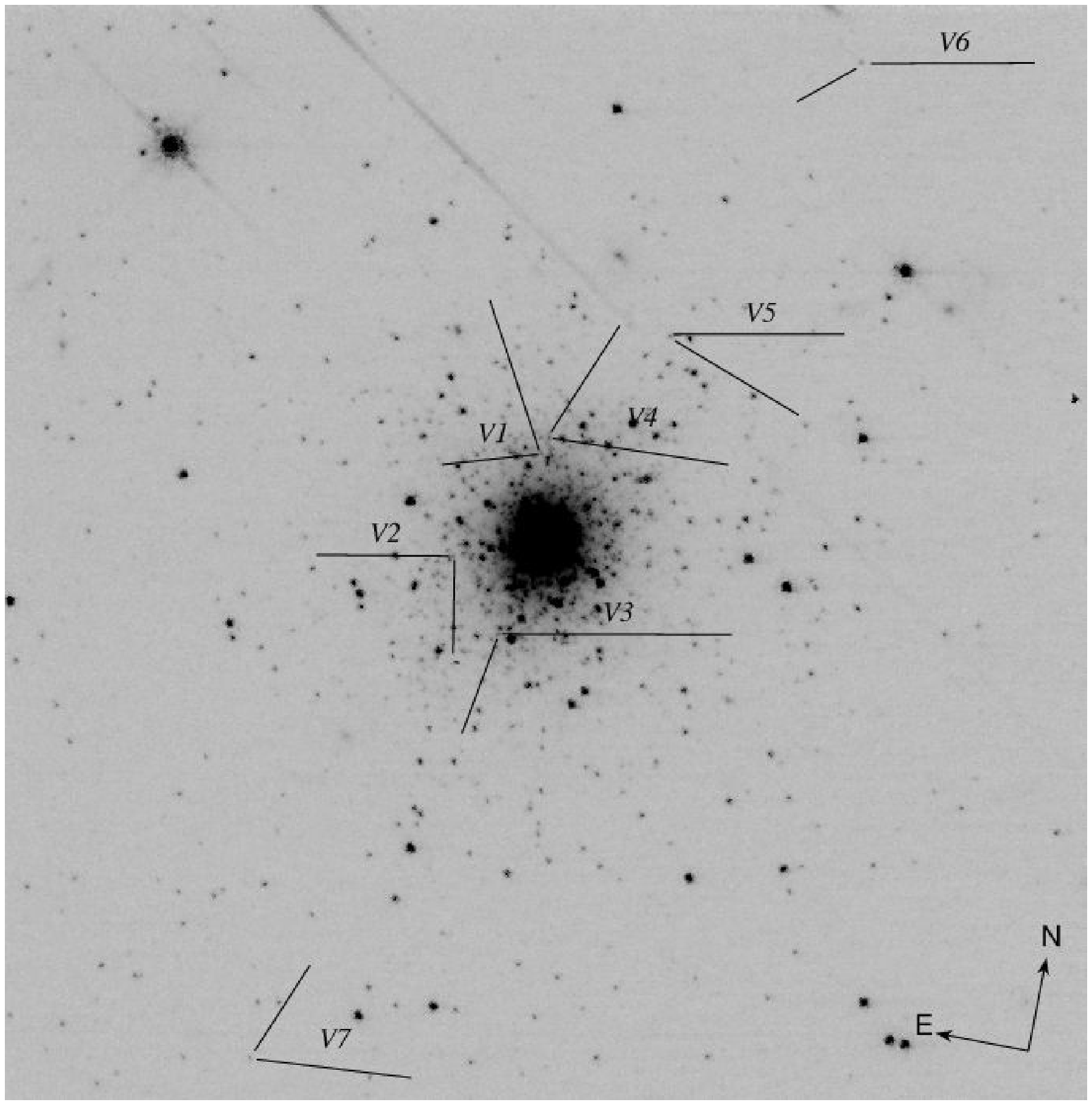} 
\caption{Finding chart of the seven variable stars identified on the $34^{\prime\prime} \times 34^{\prime\prime}$ FOV of the PC.} 
\label{pc_fchart}
\end{figure*}

%%%%%%%%%%%%%%%%%%%%%%%%%% Fig8: CMDs 3 regions
\begin{figure*} 
\centering
\includegraphics[width=16.3cm,clip]{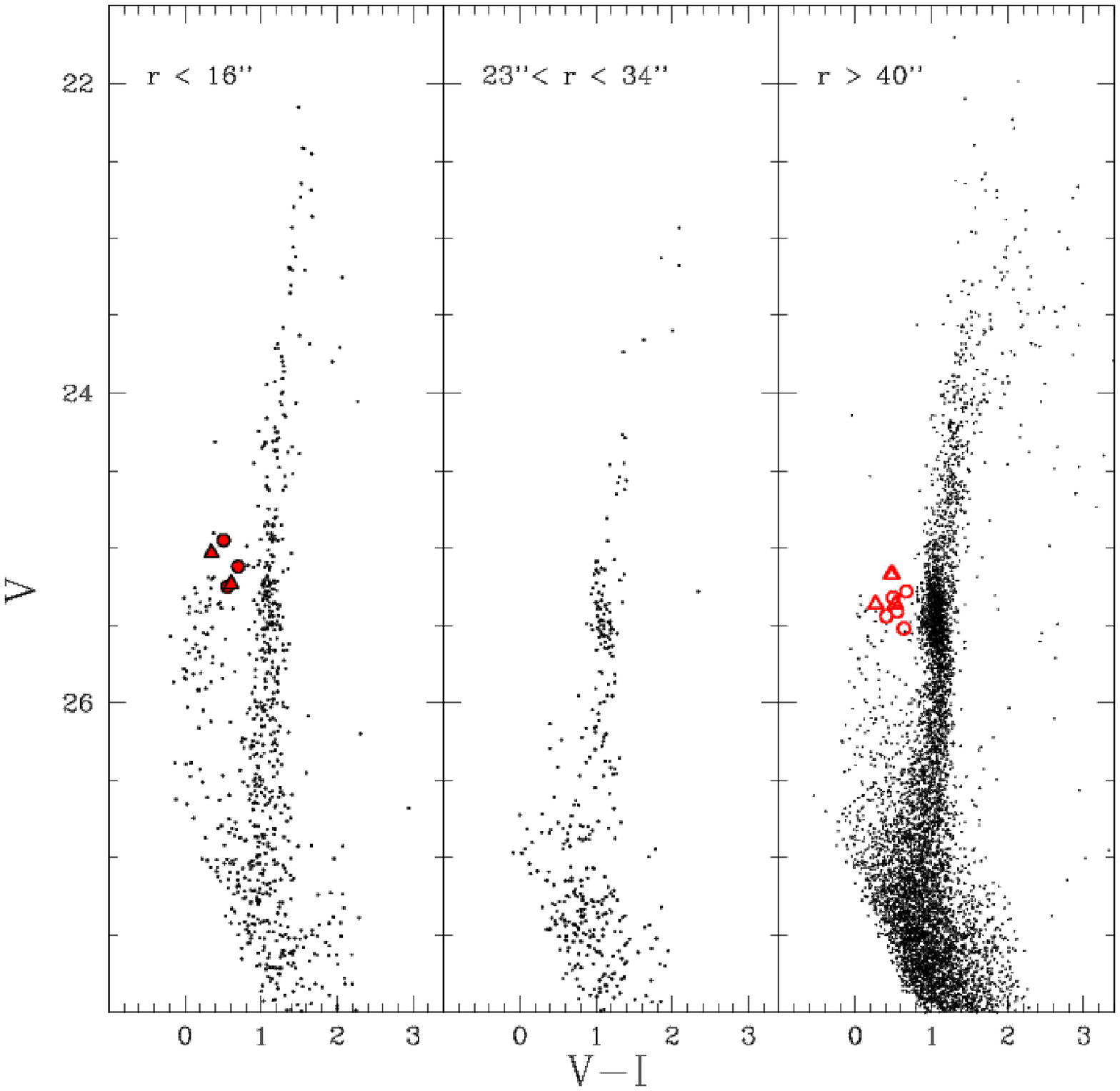} 
\caption{$V, V-I$ CMD of the sources located in 3 regions with increasing the distance 
from the cluster center.
The adopted distances from the cluster center are indicated in the panels. Symbols and color-coding are the same 
as in Figs.~\ref{pippo} and \ref{fchart}.
}
\label{cmds_f}
\end{figure*}

%%%%%%%%%%%%%%%%%%%%%%%%%% Fig9: CMDs 4 regions

\begin{figure*} 
\centering
\includegraphics[width=16.3cm,clip]{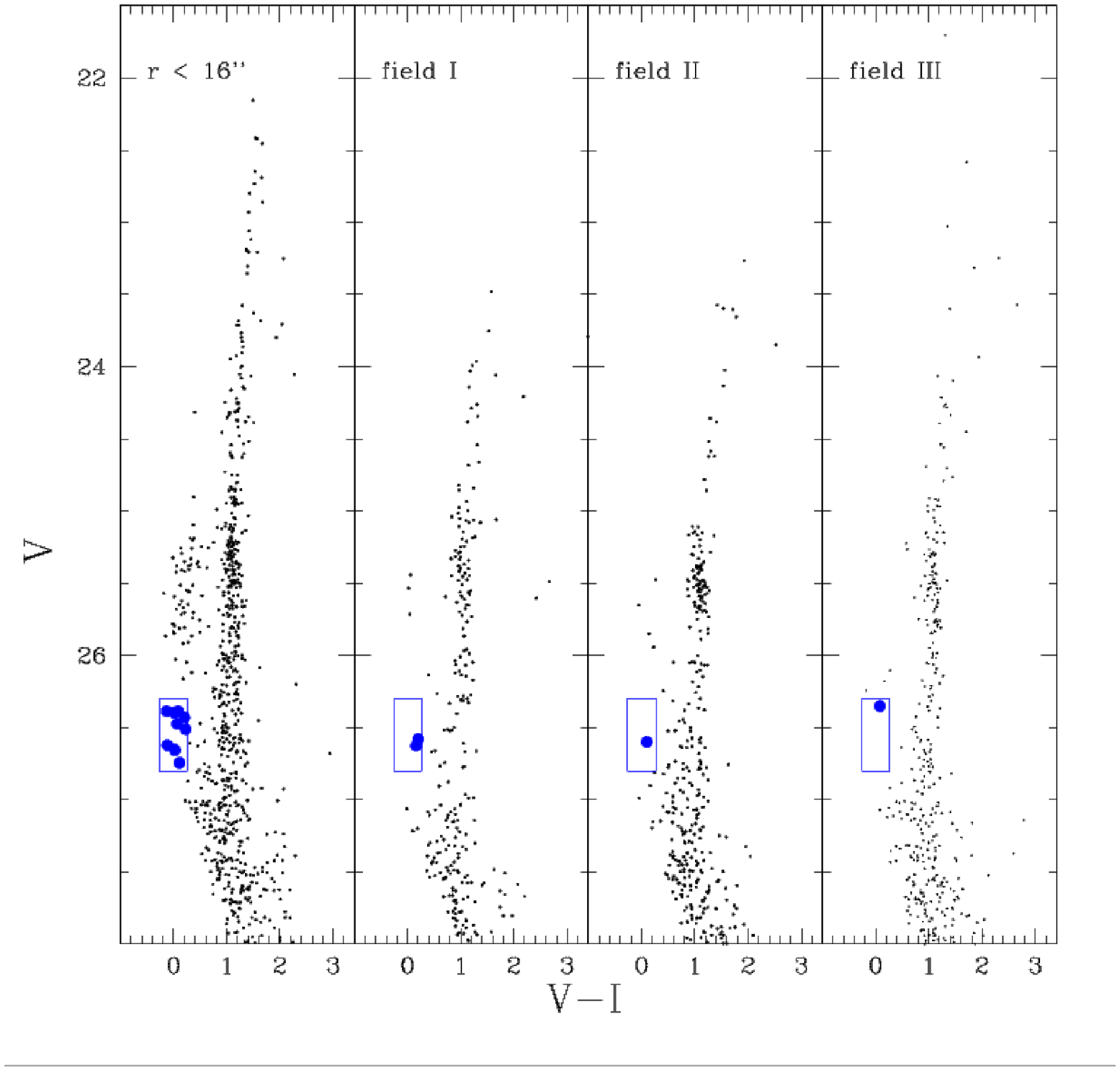} 
\caption{$V, V-I$ CMDs of the 4 regions of equal area shown in Fig.~\ref{fchart}, that were drawn with the tidal
radius of G11. We have marked with (blue) filled circles and encircled with boxes 
stars with $26.8\leq V \leq$ 26.3 mag and $V-I  \lesssim$ 0.2 mag that belong to the extreme blue HB of G11, (see text for details).
}
\label{cmds}
\end{figure*}

%%%%%%%%%%%%%%%%%%%%%%%%%% Fig10: CMD zoom
\begin{figure*} 
\centering
\includegraphics[width=16.3cm,clip]{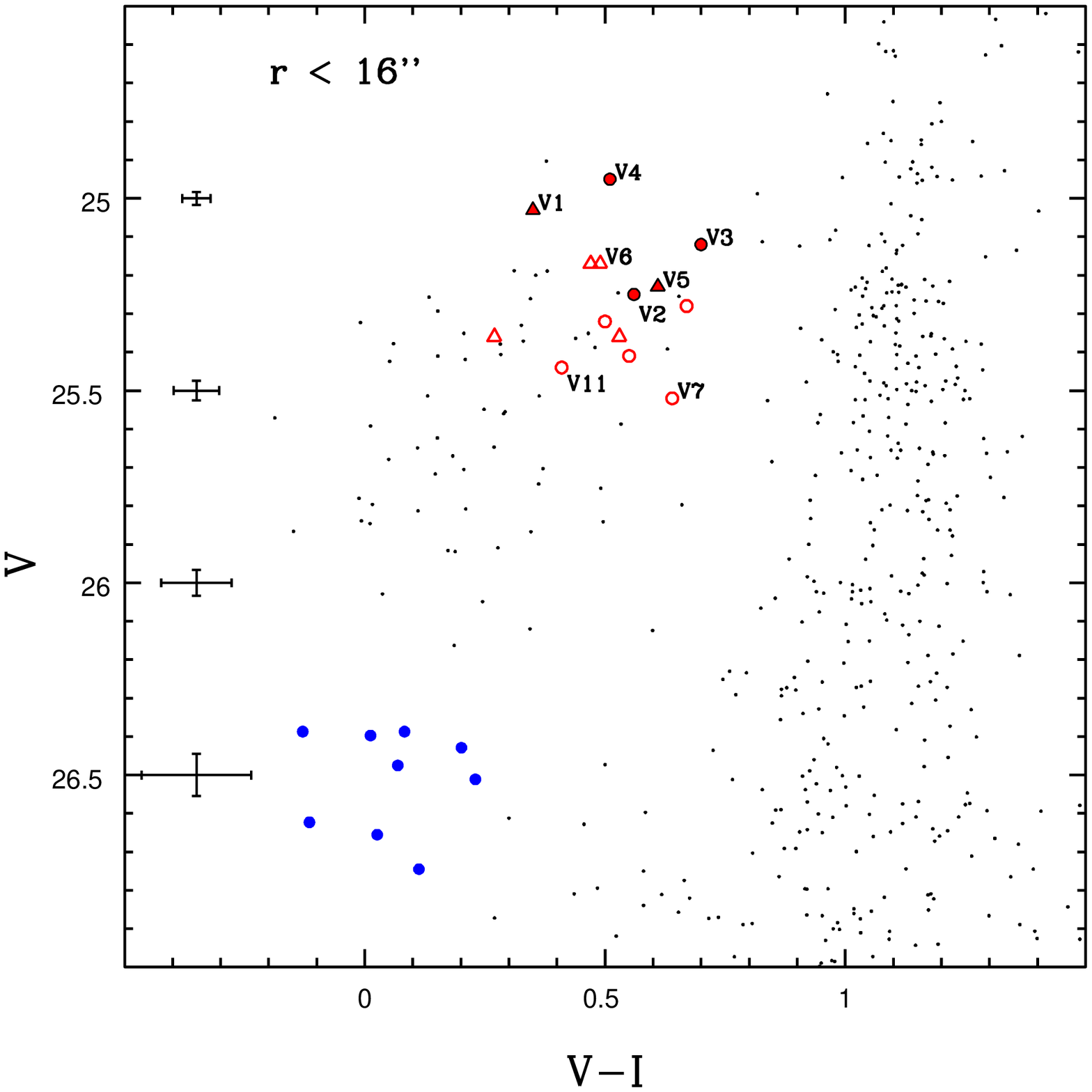} 
\caption{Blow--up on the HB for stars within 16$\arcsec$
from G11 center. Overplotted are the  RRL stars within (filled symbols) and outside (open symbols) the cluster tidal radius. Symbols and color-coding are the same as in Figs.~\ref{pippo} and \ref{fchart}. Marked with (blue) filled circles are
stars belonging to the extreme blue HB of G11.
}
\label{cmdhb}
\end{figure*}

%%%%%%%%%%%%%%%%%%%%%%%%%% Figs10: Oosterhoff plot
\begin{figure*} 
\centering
\includegraphics[width=8.7cm,clip]{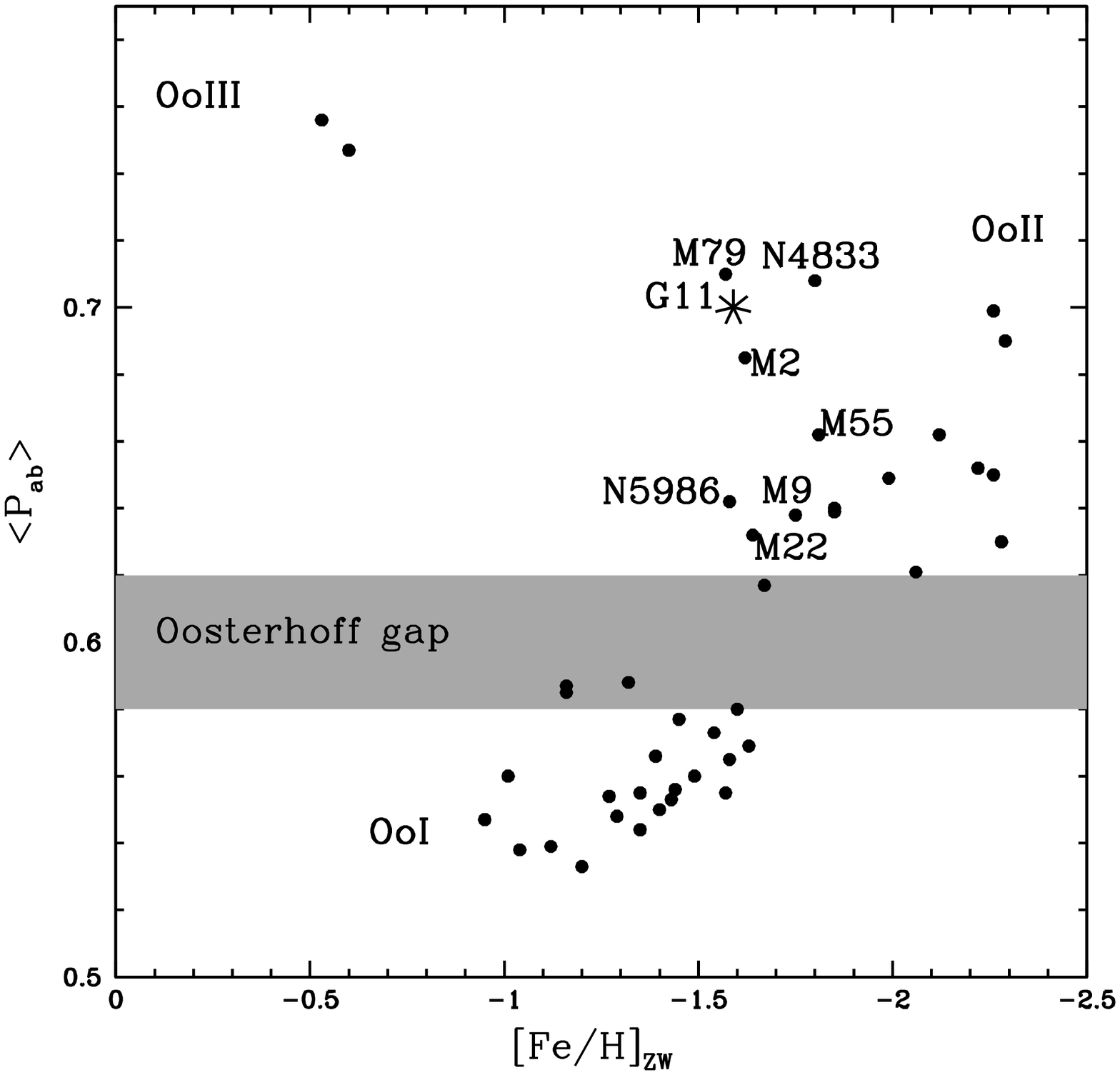} 
\includegraphics[width=8.7cm,clip]{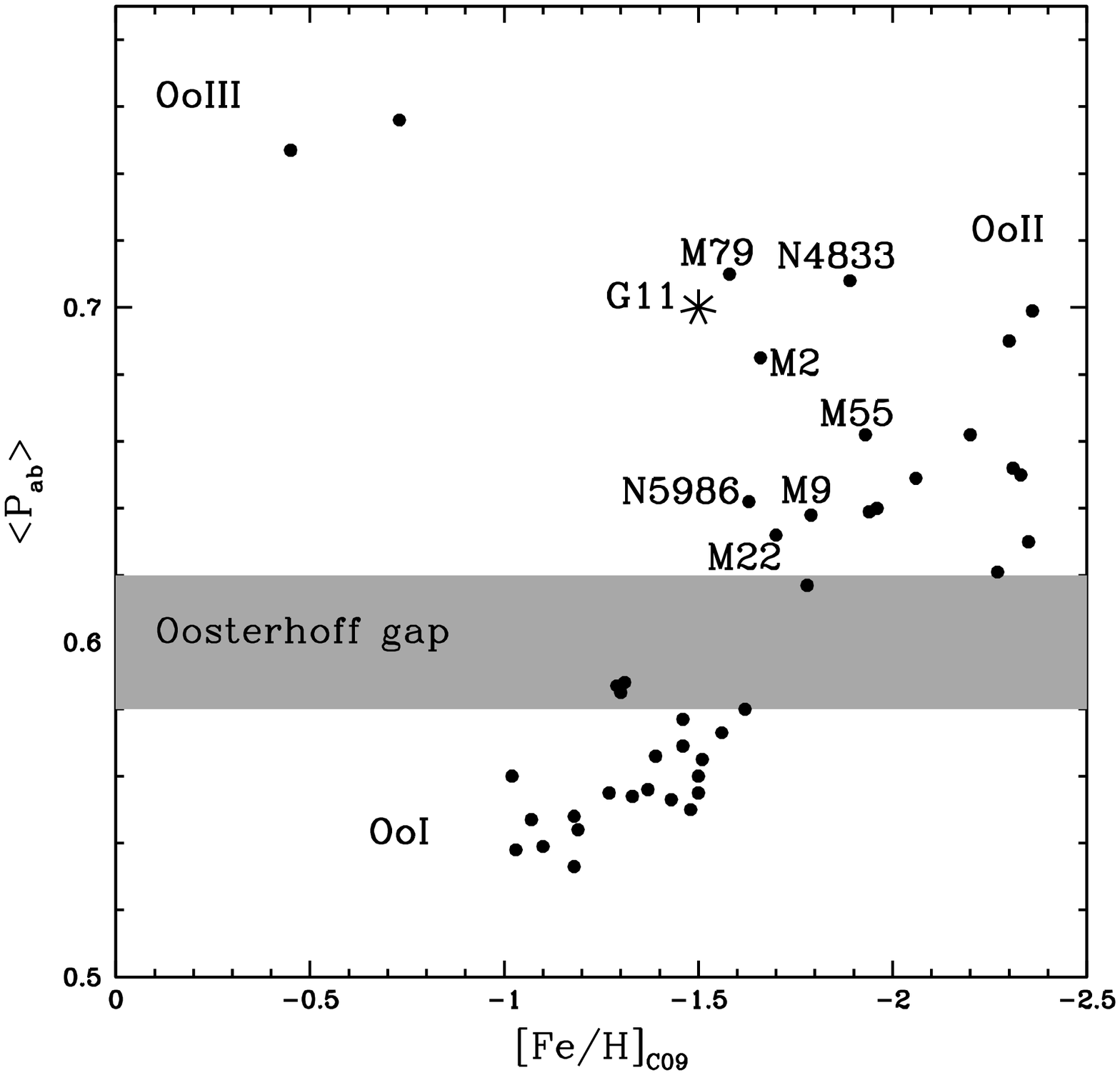} 
\caption{Mean period of the RRab stars against metallicity for the MW  GCs (dots) 
containing RRL stars, showing the well known Oosterhoff dichotomy. G11 (asterisk) 
is located in the Oo~II region.  Metallicities in the left and right panels are 
in the \cite{zinn84} and in the  \cite{carretta09} scales, respectively. 
All the GCs shown in the figure contain five or more RRab stars, except M55 which 
has only four.  The two Oo~III clusters are NGC 6388 and NGC 6441. 
%Note the reduced scatter of the Oo~I clusters when the \cite{carretta09} scale is adoped.
}
\label{oo}
\end{figure*}

\clearpage

%%%%%%%%%%%%%%%%%%%%%%%%%% Figs11: BAILEY DIAGRAMS
\begin{figure*} 
\centering
\includegraphics[width=8.7cm,clip]{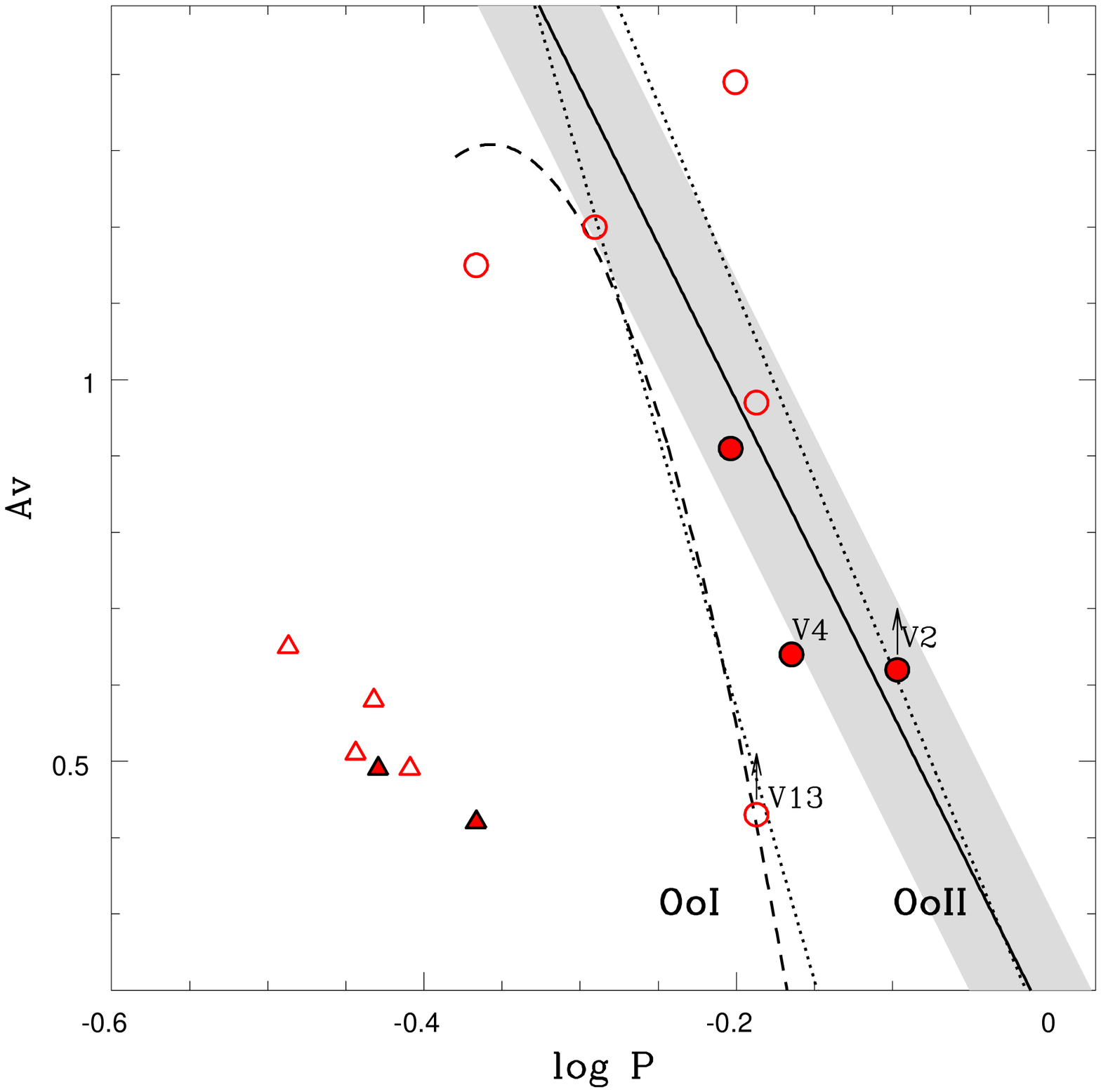} 
\includegraphics[width=8.7cm,clip]{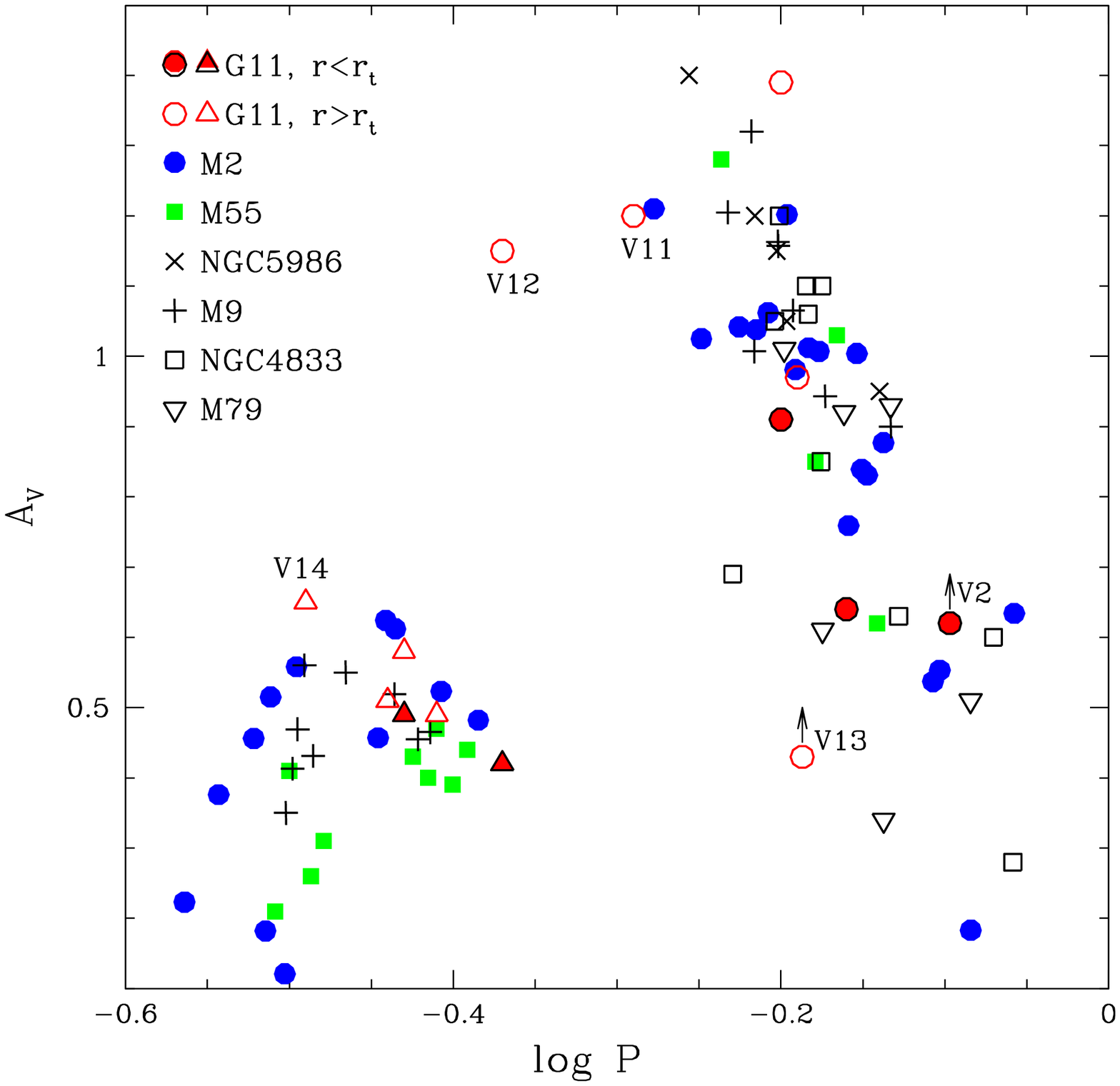} 
\caption{{\it Left panel:} $V$-band period-amplitude (Bailey) diagram for the  RRL stars identified in this study. 
Symbols and color-coding are the same as in Figs.~\ref{pippo} and \ref{fchart}. The dotted lines show the loci of the Oo~I and Oo~II GGCs according to  \citet{clement00b}.
The dashed curve represents the locus of the Oo~I GGCs  derived by \citet{cacciari05} based on
the RRL stars in M3.  The solid line shows a new locus for Oo~II GCs derived using more than 200 RRab stars in 21 Oo~II GGCs (Contreras Ramos 2012, in preparation). The 
 grey region represents the 1-$\sigma$ confidence levels of the new relation. 
{\it Right panel:} Comparison of G11 RRL stars with variables in the Oo~II GGCs M2 (\citealt{clement01}), M55 (\citealt{clement01}), NGC~5986 (\citealt{alves01}), 
M9 (\citealt{clement99}), NGC~4833 (\citealt{darragh12}), and M79 (\citealt{kains12}).
}
\label{bai}
\end{figure*}

%\clearpage

\begin{figure} 
\centering
\includegraphics[width=8.7cm,clip]{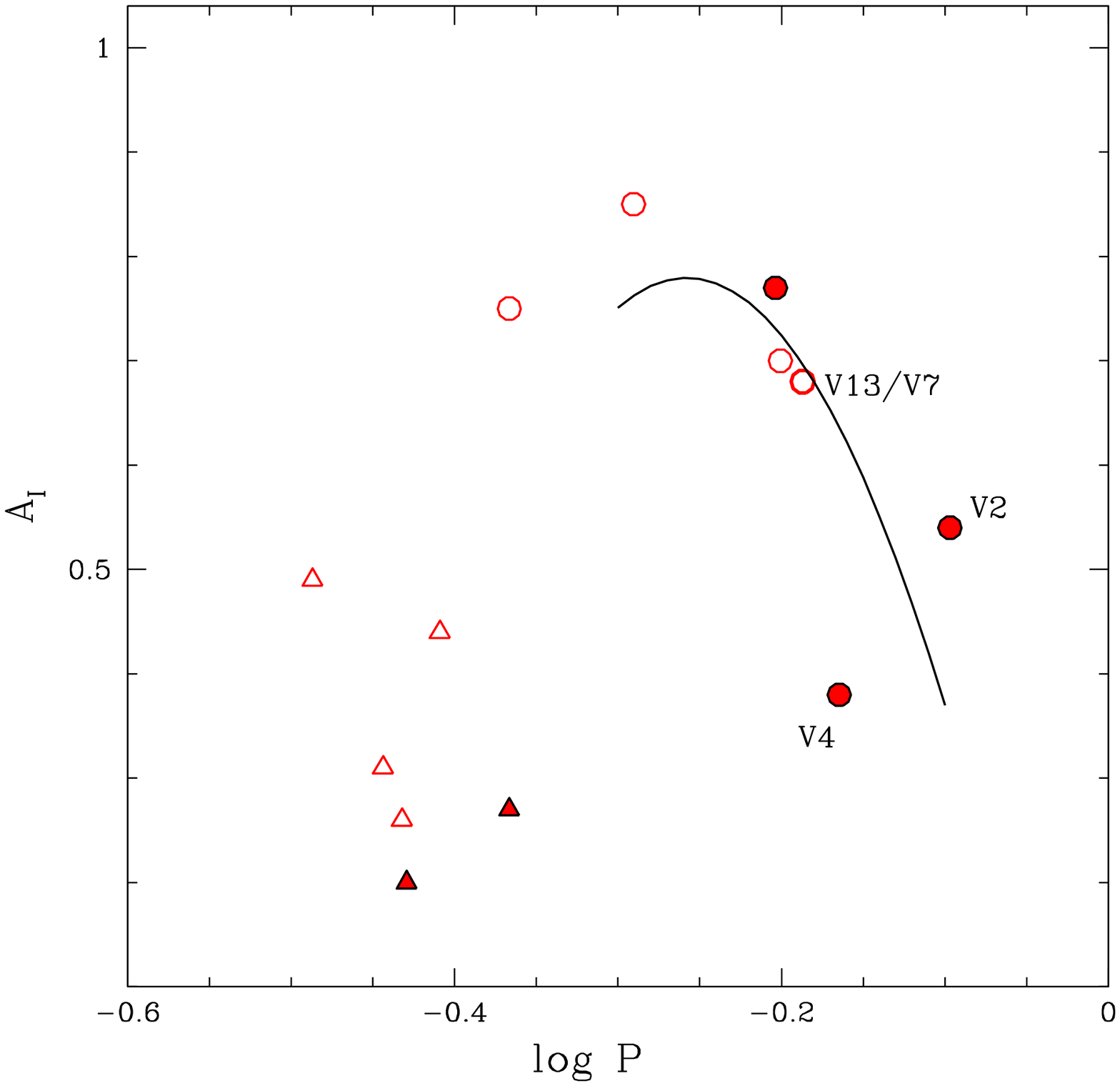} 
\caption{$I$-band period-amplitude diagram of the G11 RRL stars. 
Symbols and color-coding are the same as in Figs.~\ref{pippo}, \ref{fchart} and \ref{bai}. 
The solid curve represents the locus of the RRLs in the Oo~II GC M53 derived by \citet{arellanoferro11}.  
}
\label{baii}
\end{figure}

\clearpage

%%%%%%%%%%%%%%%%%%%%%%%%%% Fig13: CMD fit con griglia GGCs

\begin{figure*} 
\centering
\includegraphics[width=16.3cm,clip]{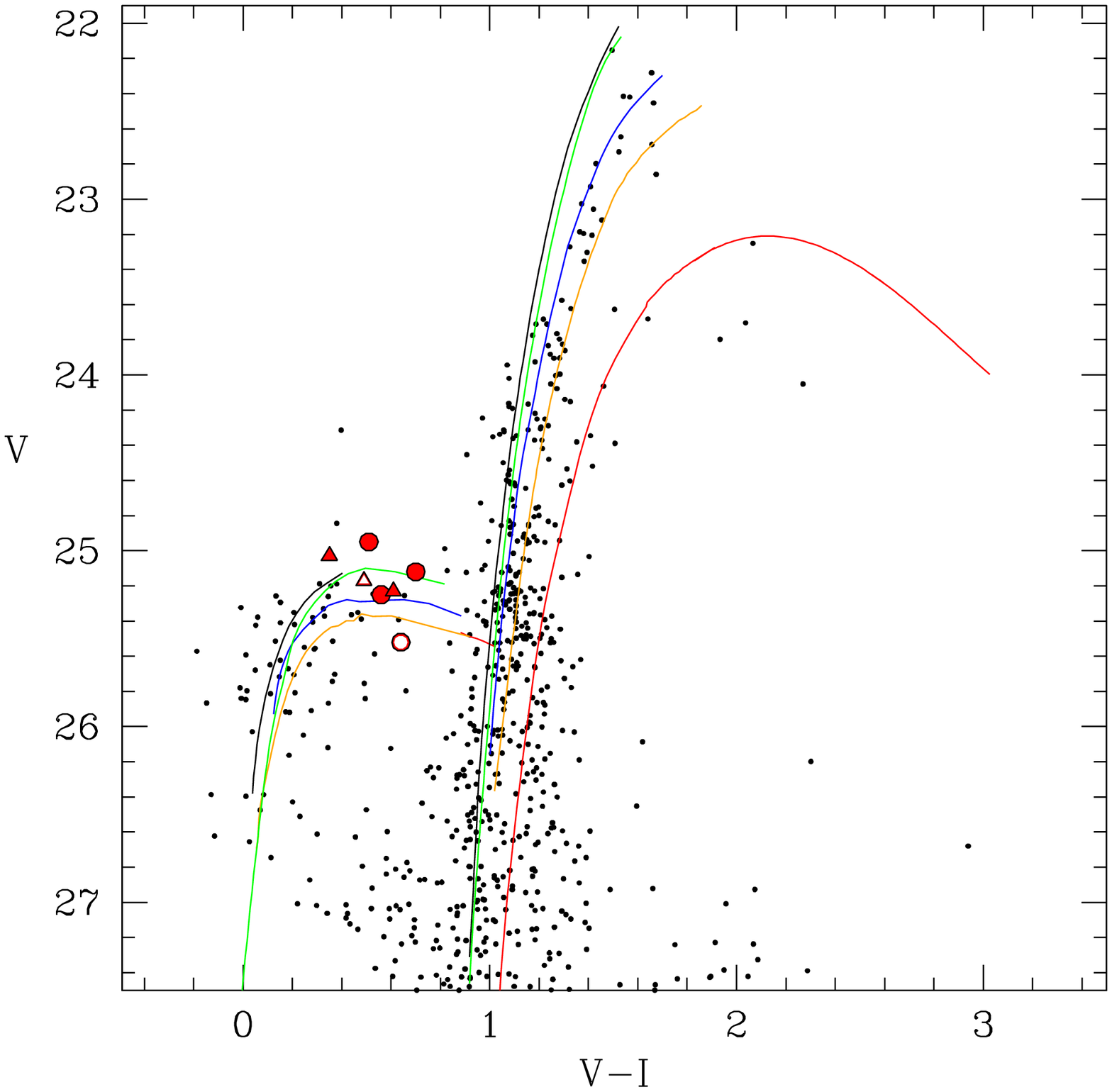} 
\caption{Fit of  the CMD,  from sources within the cluster tidal radius, to the ridgelines of the GGCs (from red to black):
47 Tuc ([Fe/H]=$-$0.71 dex); M5  ([Fe/H]=$-$1.4 dex);  M3 ([Fe/H]=$-$1.66 dex); M15 ([Fe/H]=$-$2.15 dex); and 
M92 ([Fe/H]=$-$2.24 dex).  All metallicities are on the \cite{zinn84} metallicity scale.
}
\label{fit_griglia}
\end{figure*}

\clearpage

\begin{figure*} 
\centering
\includegraphics[width=16.3cm,clip]{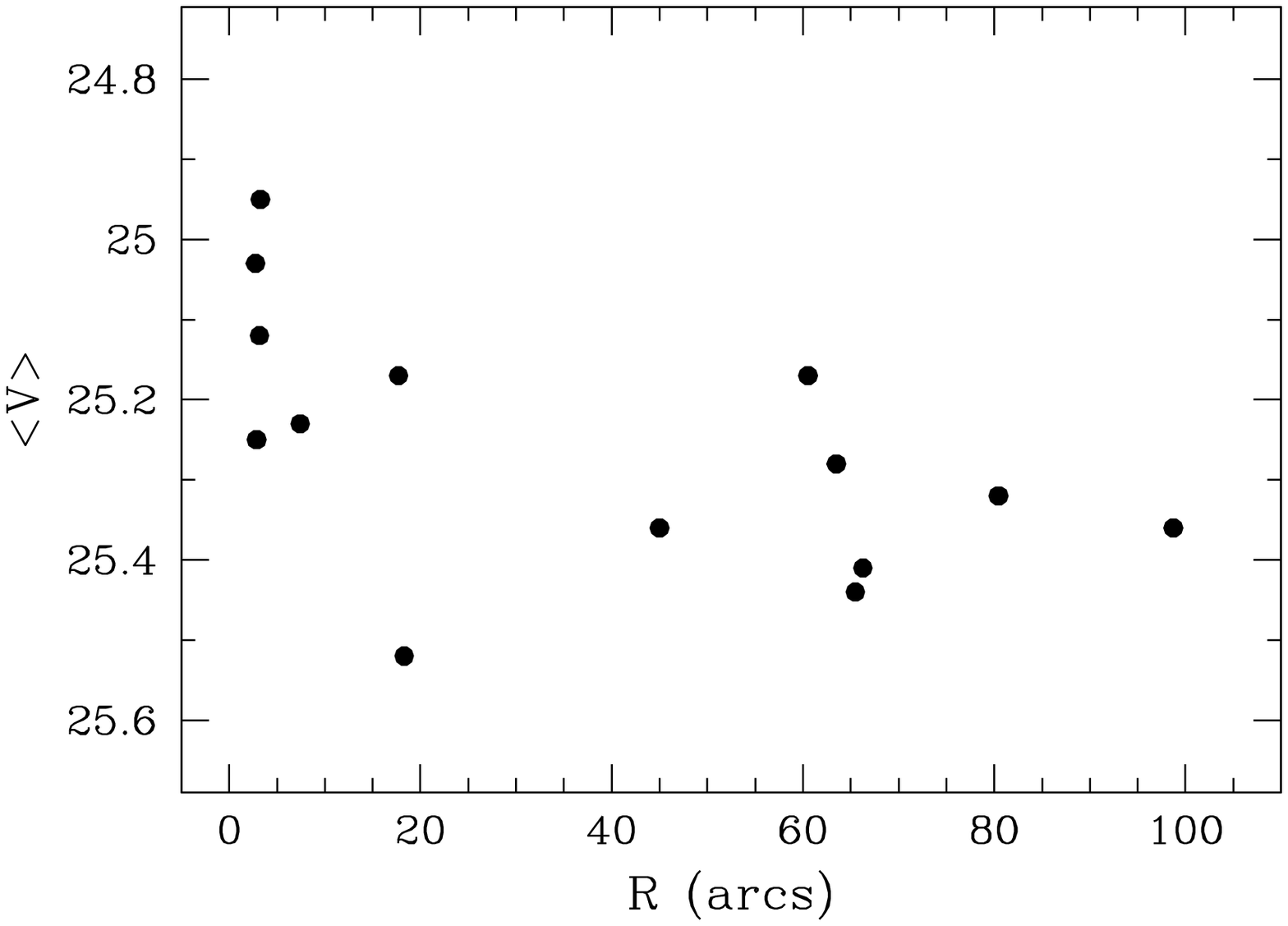} 
\caption{Run of  the mean luminosity of the individual RR Lyrae stars, 
$\langle V(RR) \rangle$, with increasing the distance from the cluster center.}
\label{vhb}
\end{figure*}

%%%%%%%%%%%%%%%%%%%%%%%%%%% TAB 00 %%%%%%%%%%%%%%%%%%%%%%%%%%%%%%%%

\begin{table*}[b]
\begin{center}
\caption[]{Basic parameters of the target M31 clusters, taken  from RBC V4.0}
\label{table01}
%	 $$
\begin{tabular}{lccccrrr}
	   \hline
	    \hline
	   \noalign{\smallskip}
{\rm ~~~~~Name }& $ V $& $(B-V)$& [Fe/H] & ${\rm [Fe/H]_{C09}}$ &{\rm X~~~~~}&{\rm Y~~~~~}&{\rm R~~~}\\    %{\rm (V-I)}&
                                & {\rm (mag)}       &    {\rm (mag)}           &  {\rm (a)}         &  {\rm (a)}     &{\rm (arcmin)} &  {\rm (arcmin)} &  {\rm (kpc) } \\ %     & 
 	    \noalign{\smallskip}
	    \hline
 	     &         &     &      &        &       &  & \\         %         &
{\rm   B293-G011} &   16.30 & 0.33& $-$1.59&$-$1.60 &   $-$61.86&  43.64 &  17.24 \\  %    0.90&
{\rm   B311-G033} &   15.45 & 0.96& $-$1.71&$-$1.73 &   $-$57.57&   0.99 &  13.11 \\  %   1.23 &
{\rm   B338-G076} &   14.25 & 0.81& $-$1.23&$-$1.20 &   $-$44.09& $-$9.05 &  10.25 \\  %   1.02&
{\rm   B343-G105} &   16.31 & 0.77& $-$1.28&$-$1.25 &   $-$57.45& $-$30.05 &  14.77 \\  %   1.04 &
{\rm   B386-G322} &   15.55 & 0.90& $-$1.09&$-$1.04 &   61.67&  $-$4.30 &  14.08 \\  %   1.16  &
{\rm   B514      }      &   15.76 &         & $-$2.06& $-$2.12 &  $-$242.32& $-$15.11 &  55.30 \\   %  1.08&
\hline
\end{tabular}
%$$
%   \tablecomments{(a) Metallicities in the C09 scale were derived from the %\citet{zinn84} 
%   values using the linear transformation provided by %\citet{carretta09}.
%   }
\end{center}
{\bf Notes.}\\
%(a) Metal abundances in column four are on the Zinn and West (1984) metallicity scale and were transformed to the new scale defined by Carretta et al. (2009)  using the linear 
(a)  [Fe/H] values are on the \citet{zinn84} metallicity scale and were transformed to the new scale defined by \citet{carretta09} (${\rm [Fe/H]_{C09}}$)  using the linear 
 transformation provided in that paper.\\
 %         	 \normalsize
\end{table*}

%%%%%%%%%%%%%%%%%%%%%%%%%%% TAB 0 %%%%%%%%%%%%%%%%%%%%%%%%%%%%%%%%

\begin{table*}[b]
 \begin{center}
      \caption[]{Log of the Observations}
	 \label{table0}
	 %\tiny
     %%%%%%%%%%%%%%%$$
	 %%%%%%%%%%%%%%%%%%%%\begin{array}{llcc}
	 $$
	\begin{array}{lccccc}
	   \hline
	    \hline
	   \noalign{\smallskip}
{\rm ~~~~~Dates}& {\rm HJD~interval}&{\rm Filter}&{\rm Exposure~length}& N &{\rm Total~ Exposure}\\
                                & (-2450000)                        &                  &                                         &            & \\
     	        &  { \rm (d)}              &            & {\rm (s)}         &  & {\rm (s)} \\
 	    \noalign{\smallskip}
	    \hline
{\rm 2000~Feb~ 3}  & 1577.73-1577.82   & F555W    & 1200/1300/2 \times1400 & 4 & 5300 \\
{\rm 2000~Feb~3 }  & 1577.86-1577.95   & F814W    & 2 \times 1300/2 \times 1400       & 4 & 5400 \\
                                &                                   &                             &                                    &    &            \\
{\rm 2007~Jul~30-31}& 4312.16-4312.82& F606W & 1100                         & 11& 12100 \\
{\rm 2007~Jul~30-31}& 4312.18-4312.84& F814W & 1100                         & 11& 12100 \\
\hline
	  %%%%%%%%%%%%%%  \end{array}
	 %%%%%%%%%%%   $$
	 \end{array}
	 $$
	 \end{center}
	 \normalsize
         \end{table*}

%%%%%%%%%%%%%%%%%%%%%%%%%%% TAB 1 %%%%%%%%%%%%%%%%%%%%%%%%%%%%%%%%

\begin{table*}[b]
\tiny
\caption[]{Identification and Properties of the RRL Stars Identified in the FOV of the G11 Observations.} 
%\centering
\begin{tabular}{llccclllllcrlcrc}
\hline
\hline
\noalign{\smallskip}
{$\rm Name$}&{\rm Camera}&{\rm x}&{\rm y}&{$\rm Dist.$}&{$\rm Type$}&~~~~P&{$\rm Epoch (max)$}&$\langle F606W\rangle$&$\langle F814W\rangle$&$ A_{F606W}$~~     &$A_{F814W}$~~&$\langle V\rangle$&$\langle I\rangle$&$ A_V$~~     &$A_I$~~\\
~~          &          &{\rm (pixel)}&{\rm (pixel)}&{\rm (arcsec)}&          & ~{\rm (days)}& ($-$2450000)   & {\rm ~~(mag)}   & {\rm ~~(mag)} & {\rm (mag)~~}   & {\rm (mag)~~}  &{\rm (mag)~}&{\rm (mag)}  &{\rm (mag)}&{\rm (mag)}\\
\noalign{\smallskip}
\hline
\noalign{\smallskip}
\multicolumn{16}{c}{\rm Variable~stars~within~ the~cluster~tidal~radius} \\
\hline
V1     	       &PC   & 346.11 & 428.77 &  2.76 & RRc   &  0.43            & ~~~4312.37  &  ~~24.93   &   ~~24.70   & ~~0.36           &  0.24       &  25.03    & 24.68  & 0.42  & 0.27 \\ %V346
V2$^{\rm a,b}$  &PC  & 414.70 & 370.00 &  2.88 & RRab &  0.79994   & ~~~1577.940 &  ~~25.11:: &   ~~24.73   & $\ge$0.54     &  0.60 & 25.25:: & 24.69 & $\ge$0.62 & 0.54 \\ %V372
V3$^{\rm a}$     &PC  & 466.85 & 400.35 &  3.16 & RRab &  0.625726 & ~~~1576.656 &  ~~24.93   &   ~~24.45   & ~~0.86           &  0.78 & 25.12 & 24.42  & 0.91 & 0.77 \\ %V363
V4$^{\rm a}$     &PC  & 335.19 & 434.09 &  3.26 & RRab &  0.6845        & ~~~1577.915 &  ~~24.80 &   ~~24.46   & ~~0.53           &  0.39 & 24.95 & 24.44  & 0.64 & 0.38 \\ %V352
V5$^{\rm a,c}$     &PC  & 267.53 & 514.53 &  7.42 & RRc    &  0.3722        & ~~~4312.760 &  ~~25.06   &   ~~24.65   & ~~0.40           &  0.21 & 25.23 & 24.62  & 0.49 & 0.20 \\%V381
\hline
\noalign{\smallskip}
\multicolumn{16}{c}{Variable~stars~beyond ~ the~cluster~tidal~radius} \\
\hline
V6$^{\rm a}$  &PC  &  84.04 & 643.00 & 17.71 & RRc    &  0.389995 & ~~~4312.2425 &  ~~25.02  &   ~~24.72   & ~~0.64           &  0.47 & 25.17 & 24.68 & 0.49 & 0.44 \\%V349 
V7     	      &PC  & 751.29 & 234.25 & 18.30 & RRab &  0.65          &  ~~~4312.750  &  ~~25.34  &   ~~24.90   & ~~0.84           &  0.57 & 25.52 & 24.88 & 0.97 & 0.68 \\ %V462
V8     	      &WF4 & 310.58 &  97.27 & 45.00 & RRc   &  0.36          & ~~~4312.248   &  ~~25.21  &   ~~24.85   & ~~0.45           &  0.32 & 25.36 & 24.83 & 0.51 & 0.31 \\ %V23306
V9$^{\rm a}$  &WF2& 532.38 & 381.04& 60.52 & RRc   &  0.36975   &  ~~~4312.172  &  ~~25.05  &   ~~24.73   & ~~0.49           &  0.25 & 25.17 & 24.70 & 0.58 & 0.26 \\ %V4365
V10                 &WF2 & 448.81 & 454.15 & 63.50 & RRab& 0.63          & ~~~4312.168   &  ~~25.10  &   ~~24.65   & ~~1.15           &  0.79 & 25.28 & 24.61 & 1.39 & 0.70 \\%V4252 
V11$^{\rm a}$&WF2 & 762.01 & 235.43 & 65.48 & RRab&  0.51226 &  ~~~1577.340  &  ~~25.31  &   ~~25.05   & ~~1.01          &  0.84 & 25.44 & 25.03 & 1.20 & 0.85 \\%V4570 
V12  	               &WF2  & 535.71 & 445.34 & 66.27 & RRab &  0.43       &  ~~~4312.37     &  ~~25.26  &   ~~24.89   & ~~1.01           &  0.76 & 25.41 & 24.86 & 1.15 & 0.75 \\%V4560 
V13$^{\rm b}$&WF4 & 518.17 & 708.96 & 80.43 & RRab &  0.65      &  ~~~4312.405    &  ~~25.20  &   ~~24.83   & ~~0.45           &  0.67 & 25.32 & 24.82 & 0.43 & 0.68 \\ %V23328
V14  	                &WF3 & 647.69 & 482.84 & 98.74 & RRc  &  0.326     &  ~~~4312.765     &  ~~25.28  &   ~~25.10   & ~~0.60           &  0.49 & 25.36 & 25.09 & 0.65 & 0.49 \\%V13676
\hline
%\enddata
\end{tabular}
\label{table2}
{\bf Notes.}\\
%\tablecomments{$^{\rm a}$ Stars with a counterpart in the 
$^{\rm a}$ Stars with a counterpart in the 
  archival WFPC2/HST data. \\
  $^{\rm b}$ Stars showing blue
  amplitudes (A$_{F606W}$ or A$_V$) smaller than the red amplitudes
  (A$_{F814W}$ or A$_I$), possibly due to contamination by companions.\\
   $^{\rm c}$ Candidate RRL star in \citet{clementini01}  that has been confirmed in the present study.
\end{table*}

%%%%%%%%%%%%%%%%%%%%%%%%%%% TAB 2 %%%%%%%%%%%%%%%%%%%%%%%%%%%%%%%%

\begin{table}[b]
\begin{center}
\caption{$F606W, F814W$ Photometry for the RRL stars Detected in this FOV of the G11 Observations. }
\vspace{0.5 cm} 
%\small
\begin{tabular}{cccc}
\hline
\hline
\multicolumn{4}{c}{G11 - Star V1 - {\rm RRab}} \\ %3996
\hline
{\rm HJD} & {\rm $F606W$}  & {\rm HJD } & {\rm $F814W$}\\ 
{\rm (--2454312)} & {\rm (mag)}  & {\rm (--2454312) } & {\rm (mag)}  \\
%            \noalign{\smallskip}
\hline
%\noalign{\smallskip}
 0.160941 &   24.83    &  0.177609 &  24.47  \\
 0.225529 &   24.92    &  0.242197 &  25.03  \\
 0.292201 &   25.01    &  0.308869 &  24.81  \\
 0.358873 &   25.19    &  0.375541 &  24.86  \\
 0.425545 &   25.09    &  0.442213 &  24.91  \\
 0.492217 &   24.91    &  0.508885 &  24.91  \\
 0.558889 &   24.87    &  0.575557 &  24.52  \\
 0.625561 &   24.97    &  0.642229 &  24.79  \\
 0.692233 &   25.03    &  0.708901 &  24.86  \\
 0.758211 &   25.29    &  0.774879 &  24.76  \\
 0.824883 &   24.93    &  0.841551 &  24.69  \\
\hline
\end{tabular}

\label{table1}
\end{center}
\medskip
(This table is available in machine-readable form in the online journal. 
A portion is shown here for guidance regarding its form and content).
\end{table}

\end{document}